\documentclass[11pt]{article}

\usepackage{amssymb, amsfonts, ifthen, epsfig}
\usepackage{graphicx}

	\usepackage{amsthm}
  	\newtheorem {theorem} {Theorem}  
  	\newtheorem {proposition} {Proposition}
  	\newtheorem {lemma} {Lemma}
  	
  	\theoremstyle{definition}
	\newtheorem{definition}{Definition}
	\newtheorem{remark}{Remark}
	
	\newtheorem{example}{Example}
	 \newcommand{\qedi}{ }

    \newcommand {\norm} [2] [] {\ensuremath{ \left\Vert  #2  \right\Vert_{#1} } }
    \newcommand {\R} {\ensuremath{\mathbb{R}}}
    
    \newcommand {\N} {\ensuremath{\mathbb{N}}}

    \newcommand {\Z} {\mathbb{Z}}

    \newcommand {\mr} {\mathrm}

    \newcommand {\C} {\mathbb C}

  \newcommand{\commut}[2]{\left[ #1 , #2 \right]}

  \newcommand{\eps}{\varepsilon}
  
  \newcommand{\tRe}{{\mathrm{Re}}}
  \newcommand{\tIm}{{\mathrm{Im}}}

\newcommand{\D}{\mathrm{d}}
\newcommand{\E}{\mathrm{e}}
\newcommand{\I}{\mathrm{i}}
\newcommand{\Or}{\mathcal{O}}

\newcommand{\floor}[1]{\left\lfloor #1 \right\rfloor}
\newcommand{\binkoeff}[2]{\left( #1 \atop #2 \right)}
\newcommand{\facnorm}[2]{\norm[(#2)]{#1}}
\newcommand{\tc}{{t_{\mathrm{c}}}}
\newcommand{\tr}{t_{\mathrm{r}}}

\newcommand{\theconstant}{\facnorm{\theta'}{1}}
\newcommand{\perturb}[1]
{
    \ifthenelse{\equal{#1}{\theta}}{\theta_{\mathrm r}}{}
    \ifthenelse{\equal{#1}{x}}{\xi}{}
    \ifthenelse{\equal{#1}{y}}{\eta}{}
    \ifthenelse{\equal{#1}{z}}{\zeta}{}
}

\newcommand{\op}{{n_\eps}}

\newcommand{\eqref}[1]{(\ref{#1})}
\newcommand{\tfrac}[2]{\textstyle \frac{#1}{#2}\displaystyle}

\newcommand{\rom}{\renewcommand{\labelenumi}{{\rm(\roman{enumi})}}}

\begin{document}

%%%%%%%%%%%%%%%%%%%%%% CMP TITELPAGE %%%%%%%%%%%%%%%%%%%%%%%%%%

%
%\title{Precise coupling terms in adiabatic quantum evolution:\\
%  The generic case.}
%\author{Volker Betz \inst{1}
%\and  Stefan Teufel \inst{2} }

%\institute{Institute for Biomathematics and Biometry, GSF
%Forschungszentrum, Postfach 1129,\\ D-85758 Oberschlei{\ss}heim,
%Germany \and Mathematics Institute, University of Warwick,
%Coventry CV4 7AL, United Kingdom
%\\[2mm]\email{volker.betz@gsf.de,
%teufel@maths.warwick.ac.uk}}
%\titlerunning{Precise coupling terms
%in adiabatic quantum evolution}

%%%%%%%%%%%%%%%%%%%%%%%% PLAIN TITLEPAGE %%%%%%%%%%%%%%%%%%%%%%%%%%%%

\title
{\Large \bf Precise coupling terms in adiabatic quantum evolution:\\
  The generic case.}
\author{Volker Betz\\ {\normalsize Institute for Biomathematics and Biometry, GSF
Forschungszentrum,}\\ {\normalsize Postfach 1129, D-85758
Oberschlei{\ss}heim,
Germany}\\{\normalsize volker.betz@gsf.de}\\[5mm] Stefan Teufel\\{\normalsize Mathematics Institute, University of
Warwick,}\\ {\normalsize Coventry CV4 7AL, United
Kingdom}\\{\normalsize teufel@maths.warwick.ac.uk}}

%%%%%%%%%%%%%%%%%%%%%%%%%%%%%%%%%%%%%%%%%%%%%%%%%%%%%%%%%%%%%%%%%%%%%%%%%%%%%%

\date{\today}

\maketitle

\begin{abstract}
For multi-level time-dependent quantum systems one can construct
superadiabatic representations in which the coupling between
separated  levels is exponentially small in the adiabatic limit.
Based on results from \cite{BeTe1} for special Hamiltonians  we
explicitly determine the asymptotic behavior of the exponentially
small coupling term for generic two-state systems with
real-symmetric Hamiltonian. The superadiabatic coupling term takes a
universal form and depends only on the location and the strength of
the complex singularities of the adiabatic coupling function.

As shown in \cite{BeTe1}, first order perturbation theory in the
superadiabatic representation then allows   to describe the
time-development of exponentially small adiabatic transitions and
thus to rigorously confirm Michael Berry's \cite{Be} predictions on
the universal form of adiabatic transition histories.

\end{abstract}

\section{Introduction and the main result}

We consider the dynamics of a two-state time-dependent quantum
system with state vector $\psi\in \C^2$ described by the
Schr\"o\-dinger equation
\begin{equation} \label{H}
\big( \I \eps \partial_t - H(t)\big)  \psi(t) = 0
 \end{equation}
 in the {\em adiabatic limit} $\eps\to 0$. The Hamiltonian $H(t)$, $t\in\R$,
 takes values in the
real-symmetric traceless $2 \times 2$-matrices of the form
\begin{equation} \label{hamiltonian}
 H(t) = {\textstyle\frac{1}{2}}\left( \begin{array}{cc} \cos\theta(t) &\,\, \,\sin\theta(t) \\
                                  \sin\theta(t) & \,-\hspace{-1pt} \cos\theta(t) \end{array}
                                  \right).
\end{equation}
This is the prototype of all adiabatic problems in quantum mechanics
and the ``Adiabatic Theorem'' states that if $H(\cdot)\in C^2(\R)$
the system \eqref{H} can be decomposed into two scalar equations
which are decoupled up to errors of order $\eps$. Under further
regularity assumptions on $H(t)$ the error bound can be improved and
for suitable real analytic $H(t)$ it is even exponentially small in
$\frac{1}{\eps}$. This adiabatic decoupling is at the basis of
understanding a large number of physical phenomena related to the
separation of time scales, ranging from the classical Stern-Gerlach
experiment for the measurement of spin to the dynamics of molecules;
see \cite{PST,Te} for recent reviews. Despite its asymptotic
smallness, the exponentially small coupling that generically remains
has itself important physical consequences such as the non-radiative
decay in molecules. In the scattering limit the exponentially small
non-adiabatic transitions are quantified by the Landau-Zener formula
and its generalizations, with rigorous justification given in
\cite{JKP,Jo}. In this work we treat the problem of explicitly
determining the exponentially small non-adiabatic coupling for
arbitrary finite times $t$ in order to obtain a complete
understanding of the nature and the time-development of
non-adiabatic transitions. Our main result is the construction of a
family of unitary maps $U^{n_\eps}_\eps(t)$ that brings \eqref{H}
into almost diagonal form \eqref{HN} with  off-diagonal elements
$c^{n_\eps}_\eps(t)$ that are exponentially small in $\eps$ and are
explicitly given at leading order.
 The construction works under
assumptions satisfied for ``generic'' Hamiltonians   in a
sense to be made precise. Our work was motivated by results of
  Berry \cite{Be} (see also  \cite{BeLi,LiBe}). He argues that the
time-development of non-adiabatic transitions is determined solely
by the complex singularities of $\theta'(t)$ closest to the real
axis, which are ``generically'' first order poles. For such generic
poles the transition histories then  have the universal form of an
error function. Despite substantial progress in  adiabatic theory
during the last decade, e.g.\ \cite{JoPf,Ne,Sj,Ma}, a rigorous
justification of Berry's conjecture for the generic case remained an
open problem until now.  We are aware only of two results
\cite{HaJo,BeTe1}, which  both  deal  with special and
``non-generic'' Hamiltonians. However, the present work is a
continuation of \cite{BeTe1} and relies on techniques and results
developed there. We also refer to \cite{BeTe1} for a more detailed
introduction and a guide to the literature on adiabatic theory in
quantum mechanics.

Before we describe our result in detail let us shortly comment on
the special form of the Hamiltonian \eqref{hamiltonian}, whose
eigenvalues are equal to $\pm\frac{1}{2}$ independent of $t$. Berry
and Lim \cite{BeLi} observed that the Schr\"odinger equation
\begin{equation} \label{tilde H}
\big( \I \eps \partial_s -\widetilde H(s)\big)  \psi(s) = 0
 \end{equation}
 for any traceless  real-symmetric Hamiltonian $\widetilde
H(s)$,
\begin{equation} \label{tilde hamiltonian}
\widetilde H(s) = \left( \begin{array}{cc}Z(s)&X(s)\\X(s)&-Z(s)
\end{array}
                                  \right)=\rho(s)\left( \begin{array}{cc} \cos\tilde \theta(s) &\,\,\, \sin\tilde\theta(s) \\
                                  \sin\tilde\theta(s) & \,-\hspace{-1pt} \cos\tilde\theta(s) \end{array}
                                  \right)\,,
\end{equation}
with eigenvalues $\rho_\pm(s)=\pm\rho(s)= \pm\sqrt{X^2(t)+Z^2(t)}$
that satisfy $\rho_+(s)-\rho_-(s)= 2\rho(s)\geq g>0$ for all
$s\in\R$, can be brought into the form \eqref{H} \&
\eqref{hamiltonian} through the invertible transformation to the
natural time variable
\begin{equation} \label{trafo}
 t(s) = 2 \int_0^s \rho(u) \, \D u\,.
 \end{equation}
While our main result is formulated for \eqref{H} and
\eqref{hamiltonian}, the difficulty of the problem stems in parts
from the fact that our assumptions must be general enough to be
satisfied by Hamiltonians \eqref{hamiltonian} that arise from
generic analytic Hamiltonians of the form \eqref{tilde hamiltonian}
through the transformation \eqref{trafo}. E.g., it turns out that
these assumptions prevent the use of standard Cauchy estimates. Note
that, as was observed by Berry \cite{Be}, Equation~\eqref{H} with a
 complex-hermitian Hamiltonian is unitarily equivalent to a similar
 equation with a
real-symmetric but $\eps$-dependent Hamiltonian. Our results on
\eqref{H} \& \eqref{hamiltonian} are sufficiently uniform to also
apply to this case and thus cover generic analytic self-adjoint
$2\!\hspace{1pt}\times\!2$-Hamiltonians, cf.\ \cite{BeTe2}.

It is well known (see \cite{BeTe1} for details and references) that
for all $n\in\N_0$ there is basis transformation such that   in the
{\em $n^{\rm th}$ superadiabatic basis} the off-diagonal elements of
the Hamiltonian are of order $\eps^{n+1}$, i.e.\ there is a unitary
$U^n_\eps(t)$ such that
\[
U^n_\eps(t)\,\big( \I \eps \partial_t -H(t)\big)\,U_\eps^{n}(t)^*\,
U^n_\eps(t) \psi(t) =: \big( \I\eps\partial_t - H^{n}_\eps(t)\big)
\psi^{n}(t) = 0
\]
with
\begin{equation}\label{HN}
H^{n}_\eps(t) = \left(\begin{array}{cc} \rho^n_\eps(t)&
c^n_\eps(t)\\[2mm]\bar c^n_\eps(t)&-\rho^n_\eps(t)\end{array}\right),\quad c^n_\eps(t)=\Or(\eps^{n+1})\,,\quad \mbox{and} \quad \psi^{n}
(t) = U_\eps^n(t) \psi(t)\,.
\end{equation}
Here $\rho^n_\eps(t) = \frac{1}{2}+\Or(\eps^2)$. Note that
$U^0_\eps(t) =:U_0(t)$ is the orthogonal transformation that
diagonalizes the symmetric matrix $H(t)$. It is independent of
$\eps$ and  maps to the  {\it adiabatic basis}. However, in general,
$\lim_{n\to\infty} |c^n_\eps(t)|=\infty$ for all $\eps>0$ and the
coupling can not be eliminated completely for fixed $\eps$ by going
to higher and higher superadiabatic bases. Instead, for each
$\eps>0$ there exists an optimal $n_\eps=n(\eps)$ for which
$n\mapsto |c^n_\eps(t)|$ attains it minimum $|c^{n_\eps}_\eps(t)|$.
This defines the {\it optimal superadiabatic basis}. In order to
determine $n_\eps$ and $c^{n_\eps}_\eps(t)$ it is necessary to
understand precisely the asymptotic behavior of the coupling
$c_\eps^n(t)$ as $n\to\infty$.

%This is the main mathematical problem solved in this paper.

As in the case of Berry's
 non-adiabatic transition histories
 %for the non-adiabatic transition histories
 one expects that for large $n$ the superadiabatic coupling function $c^n_\eps(t)$ is  determined by the
singularities of the adiabatic coupling function $c^0_\eps(t)=
\frac{\I\eps}{2}\theta'(t)$ and thus has a universal form. From the
abstract asymptotic analysis point of view this universality is just
another manifestation of Darboux' Principle, which  in its original
form says  that the late coefficients in the Taylor series of an
analytic function are determined by the convergence limiting
singularities, see Theorem~\ref{Darboux}. Dingle \cite{Di} realized
that the same idea also applies to various divergent series arising
for example from asymptotic expansions of integrals. In our case the
coefficients $c^n_\eps(t)$ are determined by a non-linear system of
recurrence relations starting with $\frac{\I}{2}\theta'(t)$ and
involving differentiation, integration and multiplication of terms.
We show that Darboux' Principle can be applied also to this system
of recurrence relations, i.e.\ that the large $n$ asymptotics  of
the coefficients $c^n_\eps(t)$ depend solely on the convergence
limiting singularities of $\theta'$. As to be discussed in
Section~2, our system of recurrence relations can be interpreted as
the formal asymptotic expansion of the solutions to a system of
ODEs. It might well be that established techniques of asymptotic
analysis can be used to determine the asymptotic behavior of
$c^n_\eps(t)$ as $n\to\infty$ by studying this system of ODEs, but
this is far from obvious. Instead, our proof is based on a direct
analysis of the recurrence relation using a family of norms tailored
to Darboux' Principle and introduced in Definition~\ref{normDef}
below. One merit of our approach is that, in principle, it can be
applied also to more complicated recurrences as arising, e.g., in
the context of constructing precise coupling terms between different
electronic levels in molecular dynamics, see \cite{BeTe3}.

 We now describe our main result in detail. Since our construction is local in
 time,
 we can restrict our attention to a compact interval
$I\subset\R$. A sufficient condition for our main theorem is that
the  singularities of the analytic continuation of $\theta'$ are of
the form
\begin{equation}\label{sing}
\theta'(z-z_0) = \frac{-\I\gamma}{z- z_0} + \sum_{j=1}^N
(z-z_0)^{-\alpha_j} h_j(z-z_0)
\end{equation}
where $|\tIm z_0|>0$, $\gamma\in \R$, $\alpha_j<1$ and   $h_j$ is
analytic in a neighborhood of $0$ for $j=1,\ldots,N$. Following
arguments of Berry and Lim \cite{BeLi} we show in Section~4 that
this condition is   fulfilled for generic Hamiltonians of the form
\eqref{tilde hamiltonian}. However, in all but a few non-generic
cases, $-\alpha_j\notin \N$ for at least one $j$.
 The
technical problem arising from this fact is that by removing the
leading singularity one does not obtain a function which is analytic
in a larger region. As a consequence, it is not sufficient to use
standard Cauchy estimates   to show that the remainder terms are
asymptotically smaller than the contribution from the leading
singularity. Instead we introduce the following norms tailored to
Darboux' Principle.

\begin{definition}\label{normDef}
Let  $\tc > 0$, $\alpha > 0$ and  $I \subset \R$ be an interval. For
$f \in C^\infty(I)$ we define
\begin{equation} \label{norm}
\facnorm{f}{I,\alpha,\tc} := \sup_{t \in I} \sup_{k \geq 0} \left|
\partial^k f(t) \right| \frac{\tc^{\alpha + k}}{\Gamma(\alpha + k)}
\leq \infty
\end{equation}
and
$$ F_{\alpha,\tc}(I) = \left \{ f \in C^{\infty}(I): \facnorm{f}{I,\alpha,\tc} < \infty \right\}.$$
\end{definition}

The connection of Definition~\ref{normDef} with \eqref{sing} is
given by Darboux' Principle, Theorem~\ref{Darboux}, which allows to
translate the information about the complex singularities of
$\theta'$ into information about the late coefficients of the Taylor
expansion of $\theta'$ on the real line. Taking $I = \{t\}$ we
obtain
\begin{equation}\label{e1}
\facnorm{f}{{\{t\}},\alpha,\tc} = C < \infty \quad
\Rightarrow\quad|f^{(k)}(t)| \leq
C\,\frac{\Gamma(\alpha+k)}{\tc^{\alpha+k}} \quad \forall k \in \N\,.
\end{equation}
Consequently $\facnorm{f}{{\{t\}},\alpha,\tc}  < \infty$ for some
$\alpha>0$ implies
 that $f$ is analytic at $t$ and that the Taylor series at
$t$ converges at least inside the disk $D_{\tc}(t)$ of radius $\tc$.
Suppose that the Taylor series has finitely many  singularities on
$\partial D_{\tc}(t)$,   all of them being of the form
$(z-z_0)^{-\alpha_k} h_k(z-z_0)$, $h_k$ analytic near the origin,
$\alpha_k>0$, then Theorem~\ref{Darboux} implies
\[
\facnorm{f}{{\{t\}},\beta,\tc}  <\infty \quad\Leftrightarrow \quad
\beta\geq \max_k\alpha_k\,.
\]

\begin{remark}
One might be tempted to think that for functions $f$ that are
analytic in $D_{\tc}(t)$ the norm $\facnorm{f}{{\{t\}},\alpha,\tc}$
is equivalent to
\[
\| f\|^{\mbox{\tiny Cauchy}}_{({\{t\}},\alpha,\tc)} :=
\sup_{|z|<\tc} |f(t+z)|\,\frac{(\tc-|z|)^\alpha}{\Gamma(\alpha)}\,.
\]
However, standard Cauchy estimates only yield
\begin{equation}\label{e2}
\| f\|^{\mbox{\tiny Cauchy}}_{({\{t\}},\alpha,\tc)}<\infty \quad
\Rightarrow\quad \exists C>0: |f^{(k)}(t)| \leq C\,\frac{\Gamma(\alpha+k+1)}{\tc^{\alpha+k}}\, \quad \forall k \in \N,
\end{equation}
which is larger than \eqref{e1} by a factor of $k+\alpha$. There may
be ways to improve, but not up to equivalence of the norms: for the
elliptic theta function $\theta_3(z) = \sum_{n=0}^\infty z^{n^2}$,
obviously $\facnorm{\theta_3}{0,\alpha,1}<\infty$ if and only if
$\alpha\geq 1$. On the other hand, an elementary estimate shows that
$\| \theta_3\|^{\mbox{\tiny Cauchy}}_{(0,\frac{1}{2},1)}<\infty$.
The reason of the discrepancy is that the Taylor coefficients of
functions with a dense set of singularities on the boundary of the
disk of convergence (as $\theta_3$ has) have worse asymptotics
than those of functions with isolated singularities. In many
problems of asymptotic analysis this lack  of preciseness of $\|
\cdot \|^{\mbox{\tiny Cauchy}}_{({\{t\}},\alpha,\tc)} $ plays no
role, since   the leading singularity  is isolated. Then one can
subtract the leading singularity  and the remainder is analytic on a
slightly larger domain. In that case Cauchy estimates applied to the
larger domain yield sufficiently small error terms. However, in our
case the form \eqref{sing} of the function near the singularity
requires the use of the precise norms $\facnorm{f}{I,\alpha,\tc}$,
since subtracting the leading singularity does not increase the
domain of analyticity.
\end{remark}

The norms $\facnorm{\cdot}{I,\alpha,\tc}$ have very convenient
mapping properties under differentiation, integration and
multiplication of functions, which are  summarized in
Proposition~\ref{key formulas 1} and Proposition~\ref{Integration}.
These mapping properties are the key ingredient for our analysis of
the recurrence relations defining $c^n_\eps(t)$, of which we will
now give precise assumptions and results. For $\gamma$, $\tr$,
$\tc\in \R$ let
$$ \theta'_0(t) = \I\,\gamma\left(\frac{1}{t - \tr + \I \tc} - \frac{1}{t -\tr- \I \tc}\right)$$
be the sum of two complex conjugate first order poles located at
$\tr\pm\I\tc$ with residues $\mp \I\gamma$.  Then, as to be
discussed
in Section 4,  for $z_0 = \tr+\I\tc$ condition \eqref{sing} generalizes to\\

\noindent{\bf Assumption 1:} {\it  On a compact interval $I\subset
[\tr-\tc,\tr+\tc]$ with $ \tr \in I$ let
 \begin{equation}
\theta'(t) = \theta'_0(t) + \theta_{\rm r}'(t)\quad\mbox{\it
with}\quad \theta_{\rm r}'(t)\in F_{\alpha,\tc}(I)
 \end{equation}
for some  $\gamma$, $\tc,\tr\in\R$, $0<\alpha<1$.}\\

It turns out that under Assumption 1 the optimal superadiabatic
basis is given as the $n_\eps^{\rm th}$ superadiabatic basis where
$0\leq \sigma_\eps <2$ is such that
\begin{equation}\label{ndef}
n_\eps = \frac{t_{\rm c}}{\eps} -1+ \sigma_\eps\qquad\mbox{is an
even integer.}
\end{equation}
Our main result is the leading order asymptotics of
$c^{n_\eps}_\eps(t)$ for $t\in I$. For times $t$ that do not belong
to an interval satisfying Assumption 1 we establish bounds on
$c^{n_\eps}_\eps(t)$ which are exponentially smaller than the
exponentially small leading order terms near the singularities. For
this we assume\\

 \noindent{\bf Assumption 2:} {\it For a compact interval $I$ and some $\tau\geq\tc$ let
 $\theta'(t) \in F_{1,\tau}(I)$.}\\

 The case of degenerate $I = \{t\}$, $t \in \R$, is explicitly allowed.
Assumptions 1 and 2 are formulated in such a way that, in principle,
$\theta'$ need only be known on the real axis. However, in practice
we will check these assumptions by analyzing the complex
singularities of the analytic continuation of $\theta'$, cf.\
Section 4.

In our main theorem we do not only control the asymptotic behavior
of the Hamiltonian in the optimal superadiabatic basis, but we also
obtain constants which are uniform on compact intervals of the other
parameters $\tc$, $\alpha$ and $\gamma$. This makes the formulation
somewhat involved, but is necessary, e.g., for the study of
hermitian but not symmetric Hamiltonians and for the study of the
scattering regime.

\begin{theorem}\label{MainTh}
Let $J_\tc\subset(0,\infty)$, $J_\alpha\subset(0,1)$ and
$J_\gamma\subset(0,\infty)$ be compact intervals.
\begin{enumerate}\rom
\item There exists   $\eps_0>0$ and a locally bounded function  $\phi_2:\R^+\to\R^+$ with $\phi_2(x)=\Or(x)$ as $x\to 0$,
  such that for all $H(t)$ as in
(\ref{hamiltonian}) satisfying Assumption~2 with $\tc\in J_\tc$, for
all  $\eps\in (0,\eps_0]$ and all $t \in I$ the elements of the
optimal superadiabatic Hamiltonian \eqref{HN} and the unitary
$U^{n_\eps}_n(t)$ with $n_\eps$ as in \eqref{ndef} satisfy
\begin{equation}\label{rhoc}
\left| \rho_\eps^{n_\eps}(t) - \frac{1}{2} \right| \leq \eps^2
\phi_2\!\left( \facnorm{\theta'}{I,1,\tau}\right) \,,\quad
\left|c^{n_\eps}_\eps(t)\right| \leq
\sqrt{\eps}\,\E^{-\frac{\tc}{\eps}(1+\ln\frac{\tau}{\tc})}
\phi_2\!\left(\facnorm{\theta'}{I,1,\tau}\right)
\end{equation}
and
\begin{equation}\label{UminusUnull}
\|U_\eps^{n_\eps}(t) - U_0(t)\| \leq \eps
\phi_2\!\left(\facnorm{\theta'}{I,1,\tau}\right).
\end{equation}\vspace{1mm}

\item Define 
$$ c_{\eps}(t) = 2\I\,\sqrt{{\textstyle\frac{2\eps}{\pi
t_{\rm c}}}}\,\sin\left({\textstyle\frac{\pi \gamma}{2}}\right)\,
\E^{-\frac{t_{\rm c}}{\eps}}\,\E^{-\frac{(t-\tr)^2}{2\eps t_{\rm
c}}} \,\cos\left({\textstyle\frac{t-\tr}{\eps}
-\frac{(t-\tr)^3}{3\eps\tc^2} + \frac{\sigma_\eps t}{\tc}} \right).$$ 
There exists   $\eps_0>0$ and a locally bounded function  $\phi_1:\R^+\to\R^+$ with $\phi_1(x)=\Or(x)$ as $x\to 0$,
  such that for all $H(t)$ as in
(\ref{hamiltonian}) satisfying Assumption~1 with $\tc\in J_\tc$,
$\alpha\in J_\alpha$, $\gamma\in J_\gamma$, for all  $\eps\in
(0,\eps_0]$ and all $t \in I$
\begin{equation}\label{Hod}
\left| c^{\op}_\eps(t) - c_{\eps}(t) 
\right| \leq \eps^{\frac{3}{2}-\alpha}\E^{-\frac{t_{\rm c}}{\eps}}
\phi_1(M),
\end{equation}
where $M= \max \left\{ \facnorm{\theta'}{I,1,\tc}, \facnorm{\perturb \theta'}{I,\alpha,\tc} \right\}$.
\end{enumerate}
\end{theorem}

\begin{remark}
Assumption 1 implies Assumption 2 on the same interval $I$ with
$\tau=\tc$. Hence, Assumption 1 implies also \eqref{rhoc} and
\eqref{UminusUnull}  with $\tau=\tc$. Furthermore this shows that in
this case the bound on $c^{n_\eps}_\eps$ in \eqref{rhoc} is optimal
with respect to the dependence on $\eps$. Again, this is only
possible since we use the precise norms $\facnorm{f}{I,\alpha,\tc}$
instead of standard Cauchy estimates.
\end{remark}

\begin{remark}
Note that the explicit term in \eqref{Hod} is  asymptotically
dominant only if $|t-\tr|=\Or(\sqrt{\eps})$. Since typically
$\tau>\tc$ in Assumption~2, for all other times $t$ the bound given
in \eqref{rhoc} is asymptotically smaller than the error term in
\eqref{Hod}.
\end{remark}

 \begin{remark}It was shown
in \cite{BeTe1} how to derive from Theorem~\ref{MainTh}, using first
order perturbation theory in the optimal superadiabatic basis,  the
universal transition histories predicted by Berry \cite{Be}.
\end{remark}

For generic analytic Hamiltonians the whole real line can be covered
by intervals satisfying either Assumption 1 or Assumption 2. Under
additional conditions on the location of the singularities of
$\theta'$, we can also consider the scattering problem and recover
the well known Landau-Zener formulas for the adiabatic transition
amplitudes. Then the decay of the exponentially small coupling for
large times can come either from $\facnorm{\theta'}{I,1,\tc}$ or from the
$\tau$-dependence of the exponent in \eqref{rhoc}.

%A detailed study of the scattering problem and further applications
%of Theorem~\ref{MainTh} to generic and non-generic physical
%Hamiltonians of the form \eqref{tilde hamiltonian} will be presented
%elsewhere \cite{BeTe2}.

Our paper consist of two parts. The main part and the key
mathematical point of our work is the proof of Theorem~\ref{MainTh}.
Our proof relies  on our previous results in \cite{BeTe1}, where we
  established Theorem~\ref{MainTh} assuming $\theta'(t) =
\theta'_0(t)$. In Section~2 we recall the necessary tools and
results from \cite{BeTe1} and prove Theorem~\ref{MainTh}, postponing
the proofs of the key inequalities to Section~3. The main
mathematical challenge is to determine the asymptotic behavior of
the solutions of a system of recurrence relation, which, as shown in
\cite{BeTe1}, yield the couplings $c^{n}_\eps(t)$. This is done in
Section~3 and the analysis heavily relies on mapping properties of
the norms $\|\cdot\|_{(I,\alpha,\tc)}$, which we believe are of
independent mathematical interest. In Section~3 we also use a
combinatorial lemma, whose  rather involved proof is postponed to
the Appendix. In Section~4 we finally discuss several issues
concerned with the transformation \eqref{trafo}. In particular we
present the argument of Berry and Lim \cite{BeLi} showing that
Assumption 1 is ``generically'' satisfied. A more detailed analysis
of this point as well as an analysis of interesting non-generic
cases and of the scattering problem are postponed to \cite{BeTe2}.
This is because the mathematical problems involved are of a
completely different
type from the main problem solved in this paper.\\

\noindent {\bf Acknowledgements:} 
%\begin{acknowledgements}
We are grateful to  Vassili
Gelfreich for several helpful remarks. We also profited from
discussions with Gero Friesecke, Alain Joye and Florian Theil. V.B.\
thanks the Mathematics Institute of the University of Warwick for
hospitality and the Symposium ``Mathematics of Quantum Systems''
organized by G.\ Friesecke for financial support.
%\end{acknowledgements}

\section{Superadiabatic representations and optimal truncation}

In this section we prove Theorem~\ref{MainTh}. The mathematical
object to control is the Hamiltonian \eqref{HN} in the
superadiabatic representation. This can be achieved by studying
superadiabatic projections. In the simple model at hand, our
understanding of these projections and their relation to the unitary
is rather complete and has been described in \cite{BeTe1}. For the
convenience of the reader, we give a synopsis here.

The $n^{\rm th}$ superadiabatic projection
\begin{equation} \label{pi n}
\pi^{(n)} = \sum_{k=0}^n \pi_k \eps^k
\end{equation}
is the unique operator (which is a $2 \times 2$ matrix in our case) with
\begin{eqnarray}
 & & (\pi^{(n)})^2 - \pi^{(n)} = \Or(\varepsilon^{n+1}) \qquad \mbox{and}  \label{pi1}\\
 & & \commut{\I\eps \partial_t - H}{\pi^{(n)}} = \Or(\varepsilon^{n+1}) \label{pi2}
\end{eqnarray}
for all $n \in \N$. Here, $\commut{A}{B}$ denotes the commutator of
the matrices $A$ and $B$; $\pi_0$ is the adiabatic projection, i.e.\
the projection onto the eigenspaces of $H$. $\pi_k$ can be
constructed recursively by using the basis
$$X = \left(\begin{array}{cc}0 & -1 \\ 1 & 0 \end{array}\right), \quad Y = -2H, \quad Z = -Y'/\theta', \quad W = 1$$
of $\R^{2 \times 2}$ and making the Ansatz
$$ \pi_k = x_kX + y_kY + z_kZ + w_k W\,.$$
It turns out that $w_k = 0$ for all $k$, while the remaining coefficients fulfill the recursive differential equations
\begin{equation}  \label{rec start}
x_1 = -\frac{\I}{2} \theta', \quad y_1 = z_1 = 0
\end{equation}
and
\begin{eqnarray}
x_n &=& -\I(z_{n-1}' - \theta' y_{n-1}), \label{xrec}\\
y_n &=& \sum_{j=1}^{n-1} (-x_j x_{n-j} + y_j y_{n-j} + z_j z_{n-j}), \label{yrec}\\
z_n &=& - \I x_{n-1}'. \label{zrec}
\end{eqnarray}
In addition, the differential equation
\begin{equation} \label{diffeq}
y_n' = - \theta' z_n
\end{equation}
holds for each $n \in \N$.

In \cite{BeTe1} we construct a unitary matrix $U^n_{\eps}(t)$ which
diagonalizes the self-adjoint matrix $\pi^{(n)}(t)$ and achieves
$$ U^n_{\eps}(t)\, \big(\I \eps \partial_t - H(t)\big) \,U_\eps^{n\,\ast}(t) = \I \eps \partial_t - \left(\begin{array}{cc} \rho^n_\eps(t)&
c^n_\eps(t)\\[2mm]\bar c^n_\eps(t)&-\rho^n_\eps(t)\end{array}\right)$$
with
\begin{equation} \label{matrix elements}
\rho_{\eps}^n(t) = 1/2 + \Or(\eps^2) \quad \mbox{and} \quad
c_{\eps}^n(t) = \eps^{n+1} (x_{n+1}(t) - z_{n+1}(t))\, (1 + \Or(\eps)).
\end{equation}
In Theorem \ref{a priori bound thm} we will prove that under Assumption 2,
\begin{equation} \label{offdiag bounds}
\left| x_{n+1}(t)\right|,\,\left|  z_{n+1}(t) \right| \leq
\frac{n!}{\tau^{n+1}} \theconstant \left(\exp( 42 \theconstant^2) -
\frac{1}{2}\right)
\end{equation}
where $\theconstant = \facnorm{\theta'}{I,1,\tau}$. Note that the
right hand side above is $\Or(\theconstant)$ as $\theconstant\to 0$
uniformly in $\tau \geq \tc\geq \inf J_{\tc} > 0$. Using
(\ref{offdiag bounds}) and the corresponding inequality for
$y_{n+1}$ from Theorem \ref{a priori bound thm} in the explicit
formulas given in Section 3 of \cite{BeTe1}, it is not difficult to
see that (\ref{UminusUnull}) holds, and that (\ref{matrix elements})
can be sharpened: there exists a locally bounded function $\phi$
with $\phi(x) = \Or(x)$ as $x \to 0$, such that for all $H(t)$ as in
(\ref{hamiltonian}) satisfying Assumption~2 with $\tc\in J_\tc$ and
for all $t \in I$
\begin{equation} \label{mat1}
\left| \rho_{\eps}^n(t) - 1/2 \right| \leq \eps^2 \phi(\theconstant)
\end{equation}
and
\begin{equation}  \label{mat2}
\left|c_{\eps}^n(t) - \eps^{n+1} (x_{n+1}(t) - z_{n+1}(t))\right| \leq  \eps^{n+2} (|x_{n+1}(t)| + |z_{n+1}(t)|) \, \phi(\theconstant).
\end{equation}
Combining (\ref{offdiag bounds}) and (\ref{mat2}), we obtain
$$ |c_{\eps}^n(t)| \leq \eps^{n+1}
\frac{n!}{\tau^{n+1}} \tilde\phi(\facnorm{\theta'}{I,1,\tau}),$$
where $\tilde\phi$ has the same properties as $\phi$, uniformly in
the class of Hamiltonians just discussed. To arrive at (\ref{rhoc}),
we take $n_\eps = \tc/\eps$, use Stirling's formula and analyze the
asymptotics of the terms involved. The procedure is performed in
detail in \cite{BeTe1}, and from the calculations there it is again
obvious that uniformity in the Hamiltonians is not lost. The only
trivial difference is that since we do not truncate at the optimal
value $\tau/\eps$ of $n$  but rather at $\tc/\eps$, we obtain in
(\ref{rhoc}) only a
 factor of $\exp\big(-\frac{\tc}{\eps} (1+\ln
\frac{\tau}{\tc})\big)$ instead of $\exp\big(-\frac{\tau}{\eps}
 \big)$. Thus we have shown part (i) of
Theorem \ref{MainTh}.

As for part (ii), let $M$ be defined as in Theorem~\ref{MainTh}. In
Theorem~\ref{perturbed situation} we will show that there exists a
locally bounded function $\phi_1$ with $\phi_1(x) = \Or(x)$ as $x
\to 0$, such that for all $H(t)$ as in (\ref{hamiltonian})
satisfying Assumption~1 with $\tc\in J_\tc$, $\alpha\in J_\alpha$,
$\gamma\in J_\gamma$, and for all $t \in I$
\begin{equation} \label{offdiag exact}
\left| x_{n+1}(t) - \I \frac{n!}{t_{\rm c}^{n+1}}
\frac{2\sin(\gamma\pi/2)}{\pi} \tRe \left( 1 + \I \frac{t-t_{\mr
r}}{t_{\rm c}} \right)^{-n-1}\right| \leq \frac{n^{-1+\alpha}
n!}{t_{\rm c}^{n+1}} \phi_1(M)
\end{equation}
provided $\theta'$ fulfills Assumption 1 and $n$ is even;
$z_{n+1}=0$ in that case. Now (\ref{mat2}), (\ref{offdiag exact})
and optimal truncation show (\ref{Hod}), and the proof of
Theorem~\ref{MainTh} is finished.

\begin{remark}
In \cite{BeTe1}, we used (\ref{diffeq}) and converted the nonlinear
recursion into the linear but nonlocal recursive
integro-differential equation
\begin{equation} \label{rec}
- z_{n+2} = z_n'' + (\theta')^2 z_n + \theta'' \int_{-\infty}^t \theta' z_n \, \D s.
\end{equation}
Since we treated the special case where $\perturb\theta' = 0$
in Assumption 1, the calculations were rather explicit and we
obtained the analogue of Theorem \ref{MainTh} with even better error
bounds. In the general situation, there is no way to avoid the
nonlinear recursion (but even so, (\ref{rec}) will be useful). As an
added bonus of not resorting to (\ref{rec}), all our results are
local.
\end{remark}

\begin{remark}
As pointed out to us by Vassili Gelfreich, (\ref{rec
start})--(\ref{zrec}) is connected to the set of singularly
perturbed algebraic-differential equations
\begin{eqnarray*}
\partial_t X(\eps,t) &=& \I \eps Z(\eps,t), \\
\partial_t Z(\eps,t) &=& \I \eps X(\eps,t) - \theta'(t) Y(\eps,t),\\
Y(\eps,t) &=& \eps(- X^2(\eps,t)+Y^2(\eps,t)  + Z^2(\eps,t))
\end{eqnarray*}
with the initial condition
$X(0,t) = \theta'(t), Y(0,t) = Z(0,t) = 0$. Indeed, consider the formal series expansion for $X$, $Y$ and $Z$, i.e.
$$ X(\eps,t) = \sum_{k=0}^{\infty} \eps^k x_{k+1},$$
with similar expressions for $Y$ and $Z$. Then (\ref{rec
start})--(\ref{zrec}) are just the equations for the coefficients of
the expansion. This opens the possibility to treat the problem using
e.g.\ Borel summation, but it is not clear to us whether this would
be successful.  On the other hand, given the connection above, it
may well be that our approach, to be presented in the following
section, can be used successfully in the theory of singularly
perturbed ODE.
\end{remark}

\section{Solving the functional recursion} \label{recursion section}

In this section we examine the recursion (\ref{rec
start})--(\ref{zrec}) and prove, in particular, the estimates
\eqref{offdiag bounds} and \eqref{offdiag exact}. The main
ingredient to our proofs is the family of norms from
Definition~\ref{normDef}. Recall that for $\tc
> 0$, $\alpha > 0$ and a compact interval $I$ we defined
\begin{equation} \label{norm2}
\facnorm{f}{I,\alpha,\tc} := \sup_{t \in I} \sup_{k \geq 0} \left| \partial^k f(t)
\right| \frac{t_c^{\alpha + k}}{\Gamma(\alpha + k)} \leq \infty
\end{equation}
and
$$ F_{\alpha,t_c}(I) = \left \{ f \in C^{\infty}(I): \facnorm{f}{I,\alpha,\tc} < \infty \right\}.$$
Often $\tc$ and $I$ will be fixed, and then we will simply write  $\facnorm{.}{\alpha}$ and $F_\alpha$.
The following mapping properties of $\facnorm{.}{\alpha}$ are crucial.

\begin{proposition} \label{key formulas 1}
Let $\tc > 0$ be fixed, $\alpha, \beta >0$ and $t \in \R$. Then
\begin{itemize}
\item[a) ]
$ \displaystyle \sup_{t \in I} \left| \partial^k f(t) \right| \leq \frac{\Gamma(\alpha+k)}{t_c^{\alpha + k}} \facnorm{f}{I,\alpha,\tc} \quad \forall k\geq 0.$
\item[b) ] $\displaystyle \facnorm{f'}{I,\alpha+1,\tc} \leq \facnorm{f}{I,\alpha,\tc}.$
\item[c) ]
Let $B(\alpha,\beta) = \Gamma(\alpha)\Gamma(\beta)/\Gamma(\alpha+\beta)$ denote the Beta function. Then
$$ \facnorm{fg}{I,\alpha+\beta,\tc} \leq B(\alpha,\beta) \facnorm{f}{I,\alpha,\tc}\facnorm{g}{I,\beta,\tc}.$$
\end{itemize}
\end{proposition}

\begin{proof} a)         and b) follow directly from the definitions.
Turning to c), for $k \geq 0$ we have
\begin{eqnarray*}
\left| (\partial^k fg)(t) \right|  & \leq & \sum_{l=0}^k \binkoeff{k}{l} \left| \partial^l f(t) \right| \left| \partial^{k-l} g(t) \right| \leq \\
& \leq & \frac{\facnorm{f}{\alpha} \facnorm{g}{\beta}}{t_c^{\alpha + \beta + k}} \sum_{l=0}^k \binkoeff{k}{l} \Gamma(\alpha+l) \Gamma(\beta+k-l).
\end{eqnarray*}
We thus have to  investigate the sum in the last line above and
relate it to $\Gamma(\alpha + \beta + k)$. To do so, we use a nice
trick, which is presumably well known. For $-1/2 < t < 1/2$ let
\begin{equation} \label{habeta}
 h_{\beta}(t) := \Gamma(\beta) \left( \frac{1}{1-t} \right)^\beta.
\end{equation}
Then $\partial^n h_{\beta} = h_{\beta + n}.$ Now consider $\partial^k (h_{\alpha} h_{\beta})$. Then on the one hand,
$$
\partial^k (h_{\alpha} h_{\beta}) = \frac{\Gamma(\alpha)\Gamma(\beta)}{\Gamma(\alpha+\beta)}
\partial^k \underbrace{\left( \Gamma(\alpha+\beta) \left(\frac{1}{1-t}\right)^{\alpha+\beta}\right)}_{= h_{\alpha+\beta}}
= B(\alpha,\beta) h_{\alpha+\beta+k}.
$$
On the other hand, of course
$$ \partial^k (h_\alpha h_\beta) = \sum_{l=0}^k \binkoeff{k}{l} \partial^l h_\alpha \partial^{k-l} h_\beta =
 \sum_{l=0}^k \binkoeff{k}{l} h_{\alpha+l} h_{\beta+k-l}.$$
Now we take $t=0$ and use $h_{\beta}(0) = \Gamma(\beta)$. Then the
above calculations give
$$ B(\alpha,\beta) \Gamma(\alpha+\beta+k) = \sum_{l=0}^k \binkoeff{k}{l} \Gamma(\alpha+l) \Gamma(\beta+k-l).$$
Inserting this in the calculations from the beginning of the proof of d), we find
$$ \left| (\partial^k fg)(t) \right| \leq \facnorm{f}{\alpha} \facnorm{g}{\beta}
\frac{\Gamma(\alpha+\beta+k)}{t_c^{\alpha + \beta + k}} B(\alpha,\beta)$$
for each $k \geq 0$, and consequently
$$ \facnorm{fg}{\alpha+\beta} \leq B(\alpha,\beta) \facnorm{f}{\alpha}\facnorm{g}{\beta}.$$ \qedi
\end{proof}

By taking $g=1$ in c) we arrive at
\begin{equation} \label{increase alpha}
\facnorm{f}{I,\alpha+\beta,\tc} \leq t_c^{\beta} \frac{\Gamma(\alpha)}{\Gamma(\alpha+\beta)} \facnorm{f}{I,\alpha,\tc}.
\end{equation}
We will also need the following somewhat more special property of the norms:
\begin{proposition} \label{Integration}
Let $s \in I$ and $\alpha > 1$.
If $f \in F_{\alpha,\tc}(I)$, then $t \mapsto \int_s^tf(r) \, \D r \in F_{\alpha-1,\tc}(I)$, and
$$ \facnorm{\int_s^t f(r) \, \D r}{\alpha-1} \leq \max \left\{ \frac{(\alpha-1) |t|}{t_c}, 1 \right\} \facnorm{f}{\alpha}.$$
In case $\alpha > 2$ and $|t-s| \leq t_c$ this simplifies to
$$ \facnorm{\int_s^t f(r) \, \D r}{\alpha-1} \leq (\alpha-1) \facnorm{f}{\alpha}.$$
\end{proposition}
\begin{proof}
We have
$$ \norm[\infty]{\int_s^t f(r) \, \D r} \leq |t-s| \norm[\infty]{f} \leq |t-s| \facnorm{f}{\alpha} \frac{\Gamma(\alpha)}{t_c^{\alpha}},$$
and for $k \geq 1$
$$\sup_{k \geq 1} \norm[\infty]{\partial^k \int_0^x f(s) \, \D s} \frac{t_c^{(\alpha-1) + k}}{\Gamma(\alpha-1 +k)} =
\sup_{k \geq 0} \norm[\infty]{\partial^k f} \frac{t_c^{\alpha + k}}{\Gamma(\alpha+k)} = \facnorm{f}{\alpha}.$$
The claim now follows from the definition of $\facnorm{\cdot}{\alpha-1}$ and the fact $\Gamma(\alpha) = (\alpha-1) \Gamma(\alpha-1)$. 
\qedi
\end{proof}

\begin{remark}
The intuition behind the norm $\facnorm{f}{I,\alpha,\tc}$ is
that when it is finite, the function $f$ behaves equally good or better
than the function $t \mapsto \frac{1}{(\I \tc + t)^\alpha}$ when
taking derivatives. The amazing and useful fact stated in
Proposition \ref{key formulas 1} c) is that multiplication not only
leaves this property intact, but even furnishes a factor that
becomes small when either $\alpha$ or $\beta$ become large. It is
this property that gets all our estimates going.
\end{remark}

\begin{theorem} \label{upper bounds}
Suppose that Assumption 2 holds, and write
$$ \theconstant := \facnorm{\theta'}{I,1,\tau} < \infty. $$
Then for each $n \in \N$,
\begin{eqnarray}
\facnorm{x_n}{I,n,\tau}  & \leq & \theconstant \left(\exp(42\theconstant^2) - \frac{1}{2}\right), \label{xub}\\
\facnorm{z_n}{I,n,\tau} & \leq & \theconstant \left(\exp(42\theconstant^2) - \frac{1}{2}\right),
\label{zub}\\
\facnorm{y_n}{I,n,\tau} & \leq & \frac{1}{n-1} \left(\exp(42\theconstant^2) - 1 \right).
\label{yub}
\end{eqnarray}
\end{theorem}

\begin{remark}
The Douglas-Adams-constant $M = 42$ comes out of our proof in a
natural way. Numerical calculations suggest that Theorem \ref{upper
bounds} holds with $M=1$, but this is probably much harder to prove.
There is also numerical evidence that the asymptotic behavior for
large $\theconstant$ is not optimal. It appears that
$\exp(M\theconstant^{3/2})$ is still an upper bound, while
$\exp(M\theconstant)$ is not.
%While unimportant for our purposes,
%obtaining sharp asymptotics for the upper bound is an interesting
%open question in its own right. (WHY?)
\end{remark}
\begin{proof}[Proof of Theorem \ref{upper bounds}]
We define $C_n$ and $D_n$ recursively through $C_1 =
\theconstant/2$, $D_1 = 0$ and
\begin{eqnarray}
C_{n} &=& \left\{
\begin{array}{lll}
    C_{n-1} & \mbox{for} & n \mbox{ even,} \\
    C_{n-1} + \frac{\theconstant}{(n-1)} D_{n-1} &\mbox{for} &  n \mbox{ odd,}
\end{array} \right. \label{CC}\\
D_{n} &=& \left\{
\begin{array}{lll}
  \sum_{k=1}^{n-1} B(k,n-k) ( C_k C_{n-k} + D_{k} D_{n-k})
  & \mbox{for} & n \mbox{ even,} \\
 0 & \mbox{for} &  n \mbox{ odd.}
\end{array} \right. \label{DD}
\end{eqnarray}
We now show that for each $n \in \N$,
\begin{eqnarray}
\facnorm{x_n}{I,n,\tau} &\leq& C_n, \label{xx}\\
\facnorm{z_n}{I,n,\tau} &\leq & C_n, \label{zz} \\
\facnorm{y_n}{I,n,\tau} &\leq& D_n \label{yy}.
\end{eqnarray}
This is checked directly for $n=1$ and $n=2$.
Suppose it holds for $n-1$. If $n$ is even, then $x_n = 0$ so (\ref{xx}) trivially holds, and (\ref{zrec}) implies
$$\facnorm{z_n}{n} = \facnorm{x'_{n-1}}{n} = \facnorm{x_{n-1}}{n-1} \leq C_{n-1} = C_{n}.$$
Proposition \ref{key formulas 1} c), (\ref{yrec}) and the fact that either $x_n= 0$ or $z_n = 0$ at any given $n \in \N$ yield
\begin{eqnarray*}
\facnorm{y_n}{n} &\leq& \sum_{k=1}^{n-1} \left( \facnorm{x_k x_{n-k}}{n}
+ \facnorm{y_k y_{n-k}}{n}
+ \facnorm{z_k z_{n-k}}{n} \right) \leq \\
&\leq & \sum_{k=1}^{n-1} B(k,n-k) (C_kC_{n-k} + D_k D_{n-k} ).
\end{eqnarray*}
If $n$ is odd, it follows from (\ref{xrec}) that $y_n = z_n = 0$, and
\begin{eqnarray*}
 \facnorm{x_n}{n} &\leq& \facnorm{z'_{n-1}}{n}
  + \facnorm{\theta' y_{n-1}}{n}  \leq \\
 &\leq& \facnorm{z_{n-1}}{n-1} + \frac{1}{n-1}\facnorm{\theta'}{1}  \facnorm{y_{n-1}}{n-1} \leq C_{n-1} + \frac{\theconstant}{n-1} D_{n-1}.
 \end{eqnarray*}
This proves (\ref{xx})--(\ref{yy}). From (\ref{CC}) it follows
immediately that
\begin{eqnarray}
C_n &=& C_{n-1} + \frac{\theconstant}{n-1}{D_{n-1}} = C_{n-2} + \frac{\theconstant}{n-1}{D_{n-1}} + \frac{\theconstant}{n-2}{D_{n-2}} = \nonumber\\
&=& \ldots = C_1 + \theconstant \sum_{j=2}^{n-1} \frac{D_j}{j} = \theconstant\left(\frac{1}{2} + \sum_{j=2}^{n-1} \frac{D_j}{j} \right). \label{C from D}
\end{eqnarray}
Thus it is sufficient to control the $D_j$. We claim:
\begin{lemma} \label{CD bound}
For $M \geq 42$ and all even $n \in \N$,
\begin{equation} \label{D estimate}
 D_n \leq \frac{1}{n-1} \sum_{j=1}^{n/2} \frac{\theconstant^{2j}M^j}{j!}.
 \end{equation}
\end{lemma}
The proof of this purely combinatorial fact is somewhat involved and
deferred to the appendix. From (\ref{D estimate}) and (\ref{yy}) it
now follows immediately that
$$\facnorm{y_n}{n} \leq D_n \leq \frac{\exp(M \theconstant^2) - 1}{n-1}$$
for all $n \in \N$. Using (\ref{xx}) and (\ref{C from D}) we obtain
\begin{eqnarray*}
\facnorm{x_n}{n} & \leq & C_n \leq \theconstant \left( \frac{1}{2} + (\exp(M \theconstant^2) - 1) \sum_{j=2}^{n-1} \frac{1}{j(j-1)} \right)\leq \\
&\leq& \theconstant \left(\exp(M \theconstant^2) - \frac{1}{2}\right),
\end{eqnarray*}
and the same estimate applies to $\facnorm{z_n}{n}$. The proof is finished. \qedi
\end{proof}

Proposition \ref{key formulas 1} a) now immediately implies

\begin{theorem} \label{a priori bound thm}
Suppose that Assumption 2 holds and write
$$ \theconstant = \facnorm{\theta'}{I,1,\tau}.$$
Then for each $n \in \N$ and each $t \in \R$, we have
\begin{eqnarray*}
\sup_{t \in I}|x_n(t)| & \leq & \frac{(n-1)!}{\tau^n} \theconstant \left(\exp( 42 \theconstant^2) - \frac{1}{2}\right), \\
\sup_{t \in I}|z_n(t)| & \leq & \frac{(n-1)!}{\tau^n} \theconstant \left(\exp( 42 \theconstant^2) - \frac{1}{2}\right), \\
\sup_{t \in I}|y_n(t)| & \leq & \frac{(n-2)!}{\tau^n} \left(\exp(42 \theconstant^2) - 1 \right).
\end{eqnarray*}
\end{theorem}

 It is interesting
that although the inequalities in Theorem \ref{a priori bound thm}
were derived using some seemingly rather crude estimates, they are
optimal up to constants in the two most important asymptotic
regimes: For large $n$ and each $\theconstant$ as well as for small
$\theconstant$ and each $n$ the results in \cite{BeTe1} are an
example that displays exactly the asymptotic behavior predicted by
Theorem \ref{a priori bound thm}.

Surprisingly, the accurate asymptotics of the  recursion (\ref{rec
start})--(\ref{zrec}) under Assumption 1 are not difficult to obtain
from the results of \cite{BeTe1} once we have the uniform bounds of
Theorem \ref{a priori bound thm} and use our norms.

\begin{theorem} \label{perturbed situation}
Let $J_\tc\subset(0,\infty)$, $J_\alpha\subset(0,1)$ and
$J_\gamma\subset(0,\infty)$ be compact intervals. There exists a
locally bounded function  $\phi_1:\R^+\to\R^+$ with
$\phi_1(x)=\Or(x)$ as $x\to 0$, such that for all $H(t)$ as in
(\ref{hamiltonian}) satisfying Assumption~1 with $\tc\in J_\tc$,
$\alpha\in J_\alpha$, $\gamma\in J_\gamma$ and all $t \in I$ we have
\begin{eqnarray}
&& \hspace{-1cm} \left| x_n(t)  -  \I c_{\gamma} \frac{(n-1)!}{t_{\rm c}^n} \tRe
\left(\left( 1 - \I \frac{t-t_{\mr r}}{t_{\rm c}} \right)^{-n}\right)\right| \leq  \frac{(n-1)!}{t_{\rm c}^n} n^{-1+\alpha} \phi_1(M) , \label{xpert}\\
%%&& |y_n(t)| \leq n^{-1+\alpha} \frac{(n-1)!}{t_{\rm c}^n}  \phi(M), \label{ypert}\\
&& \hspace{-1cm} \left| z_n(t) + c_{\gamma} \frac{(n-1)!}{t_{\rm c}^n}
\tIm \left(\left( 1 - \I
\frac{t- t_{\mr r}}{t_{\rm c}} \right)^{-n}\right)\right| \leq \frac{(n-1)!}{t_{\rm c}^n} n^{-1+\alpha} \phi_1(M), \label{zpert}
\end{eqnarray}
where $M= \max \left\{ \facnorm{\theta'}{I,1,\tc}, \facnorm{\perturb \theta'}{I,\alpha,\tc} \right\}$ and $c_{\gamma} = \frac{2\sin(\gamma\pi/2)}{\pi}$.
\end{theorem}

\begin{proof}
Let $x_{n,0}$, $y_{n,0}$ and $z_{n,0}$ be defined via the recursion
(\ref{xrec}) - (\ref{zrec}) started with $x_{1,0} = \I \theta'_0/2$.
This is the situation where $\theta'$ just consists of a pair of
simple poles, and Theorem~3 of \cite{BeTe1} implies (\ref{xpert})
and (\ref{zpert}) for $x_{n,0}$, $y_{n,0}$ and $z_{n,0}$ and any
$\alpha < 1$. Uniformity in the parameters $\gamma$ and $\tc$ is not
spelled out there, but again it is easy to derive from the estimates
given. Let us now write
$$ \perturb{x}_n = x_n - x_{n,0},
\quad \perturb{y}_n = y_n - y_{n,0}
\quad \mbox{and} \quad \perturb{z}_n = z_n - z_{n,0}.$$
The proof will be done as soon as we show
\begin{equation}
\left| \perturb{x}_n \right|, \left| \perturb{z}_n \right|  \leq \frac{(n-1)!}{t_{\rm c}^n} n^{-1+\alpha} \phi_1(M)
\end{equation}
for some $\phi$ with the properties given in the Theorem, uniformly in the parameters $\alpha$ and $\tc$.
Without loss we assume $\tr = 0$, and we write $F_k$ instead of $F_{k,\tc}(I)$ etc. The main step is
\begin{lemma} \label{bounded sequence}
$\perturb{x}_n \in F_{n-1+\alpha}$ and $\perturb{z}_n \in F_{n-1+\alpha}$ for each $n \in \N$, and there exists $\phi: \R^+ \to \R^+$ with $\phi(x) = \Or(x)$ as $x \to 0$ such that
$$ \facnorm{\perturb{z}_n}{n-1+\alpha} \leq \phi(M) \quad \mbox{and} \quad
\facnorm{\perturb{x}_n}{n-1+\alpha} \leq \phi(M) $$
for all $n \in \N$, uniformly in $\alpha \in J_{\alpha}$, $\tc \in J_{\tc}$.
\end{lemma}
\begin{proof}[Proof of Lemma \ref{bounded sequence}]
We first prove the assertion for the $\perturb{z}_n$. For odd $n$, $\perturb{z}_n = 0$ and there is nothing to prove, so let $n$ be even.
Since $\perturb{z}_2 = \perturb\theta''/2$, the assertion is true for $n=2$.
Using (\ref{diffeq}) along with the recursion, we find
\begin{equation} \label{rec2}
- z_{n+2} = z_n'' + (\theta')^2 z_n + \theta'' \left( \int_{0}^t \theta' z_n \, \D s + y_n(0) \right),
\end{equation}
which is obviously just another way to write (\ref{rec}).
We decompose (\ref{rec2}) into terms which contribute to $z_{n+2,0}$ and those that do not, with the result
\begin{eqnarray}
&&\hspace{-7mm}z_{n+2}
%&=& z_n'' + (\theta')^2 z_n + \theta'' \int_0^t \theta' z_n \, ds + \theta'' y_n(0) = \nonumber \\
= \underbrace{z_{n,0}'' + (\theta'_0)^2 z_{n,0} +
\theta''_0 \left( \int_0^t \theta'_0 z_{n,0} \, ds + y_{n,0}(0) \right)}_{= z_{n+2,0}} + \label{zeile1} \\
&& \hspace{-7mm} + \perturb{z}_n'' + (\theta')^2 \perturb{z}_n +  (2 \theta'_0 \perturb\theta' + (\perturb\theta')^2) z_{n,0} + \label{zeile2} \\
&& \hspace{-7mm} + \theta'' \int_0^t \theta' \perturb{z}_n \, ds +
 \theta''_0 \int_0^t \perturb\theta' z_{n,0} \, ds + \perturb\theta'' \int_0^t \theta' z_{n,0} \, ds + \theta'' y_n(0) - \theta''_0 y_{n,0}(0). \label{zeile3}
\end{eqnarray}
The terms in (\ref{zeile2}) and (\ref{zeile3}) contribute to
$\perturb{z}_{n+2}$, and we are going to estimate the
$\facnorm{.}{n+1+\alpha}$-norm of each of them, using Propositions
\ref{key formulas 1} and \ref{Integration}. Starting with
(\ref{zeile2}), we have
\begin{eqnarray}
\facnorm{\perturb{z}''}{n+1+\alpha} &\leq& \facnorm{\perturb{z}}{n-1+\alpha}, \label{est1}\\
\facnorm{(\theta')^2\perturb{z}_n}{n+1+\alpha} &\leq& B(2,n-1+\alpha) \facnorm{(\theta')^2}{2} \facnorm{\perturb{z}}{n-1+\alpha} \leq \nonumber \\
&\leq& \frac{1}{(n+\alpha)(n-1+\alpha)} \facnorm{\theta'}{1}^2 \facnorm{\perturb{z}}{n-1+\alpha}, \label{est2}
\end{eqnarray}
and
\begin{eqnarray}
 \facnorm{(2 \theta'_0 \perturb\theta' + (\perturb\theta')^2) z_{n,0}}{n+1+\alpha}
  &\leq& B(1+\alpha,n) \facnorm{\perturb\theta' (\theta'_0 + \theta')}{1+\alpha} \facnorm{z_{n,0}}{n} \nonumber \\
 && \hspace{-3cm}\leq  \frac{\Gamma(\alpha)\Gamma(n)}{\Gamma(n+\alpha+1)} \facnorm{\perturb\theta'}{\alpha}\left(\facnorm{\theta'_0}{1}
 + \facnorm{\theta'}{1}\right) \facnorm{z_{n,0}}{n}. \label{est3}
\end{eqnarray}
Turning to (\ref{zeile3}), let us first note that
$$ \facnorm{\int_0^t \theta' \perturb{z}_n \, ds}{n-1+\alpha} \leq (n-1+\alpha) \facnorm{\theta' \perturb{z}_n}{n+\alpha} \leq
\facnorm{\theta'}{1} \facnorm{\perturb{z}_n}{n-1+\alpha}.$$
Similarly,
\begin{eqnarray*}
 \facnorm{\int_0^t \perturb\theta' z_{n,0} \, ds}{n-1+\alpha} &\leq& (n-1+\alpha) B(\alpha,n) \facnorm{\perturb\theta'}{\alpha} \facnorm{z_{n,0}}{n} = \\
  &=& \frac{\Gamma(\alpha)\Gamma(n)}{\Gamma(n-1+\alpha)}\facnorm{\perturb\theta'}{\alpha} \facnorm{z_{n,0}}{n}
 \end{eqnarray*}
and
$$ \facnorm{\int_0^t \theta' z_{n,0} \, ds}{n} \leq \facnorm{\theta'}{1}\facnorm{z_{n,0}}{n}.$$
With these estimates we obtain
\begin{eqnarray}
\facnorm{ \theta'' \int_0^t \theta' \perturb{z}_n \, ds}{n+1+\alpha} & \leq & \frac{1}{(n+\alpha)(n+\alpha-1)} \facnorm{\theta'}{1}^2
 \facnorm{\perturb{z}_n}{n-1+\alpha}, \label{est4}\\
\facnorm{\theta''_0 \int_0^t \perturb\theta' z_{n,0} \, ds}{n+1+\alpha} &\leq& \frac{\Gamma(\alpha)\Gamma(n)}{\Gamma(n+1+\alpha)}
\facnorm{\theta'_0}{1}\facnorm{\perturb\theta'}{\alpha} \facnorm{z_{n,0}}{n}, \label{est5}\\
\facnorm{\perturb\theta'' \int_0^t \theta' z_{n,0} \, ds}{n+1+\alpha} &\leq& B(1+\alpha,n) \facnorm{\perturb\theta'}{\alpha}
\facnorm{\theta'}{1}\facnorm{z_{n,0}}{n}. \label{est6}
\end{eqnarray}
Finally,
\begin{eqnarray} 
\facnorm{\theta''y_n(0)}{n+1+\alpha}
%=  |y_n(0)| \facnorm{\theta''}{n+1+\alpha}
&\leq& |y_n(0)| \frac{\tc^{n-1+\alpha}}{\Gamma(n+1+\alpha)}\facnorm{\theta''}{2} \nonumber\\
&\leq&  \tc^{-1+\alpha} \frac{\Gamma(n)}{\Gamma(n+1+\alpha)}
 \facnorm{\theta'}{1} \facnorm{y_n}{n}, \label{est7}
\end{eqnarray}
and
\begin{equation} \label{est8}
 \facnorm{\theta''_0 y_{n,0}(0)}{n+1+\alpha} \leq \tc^{-1+\alpha} \frac{\Gamma(n) }{\Gamma(n+1+\alpha)} \facnorm{\theta'_0}{1} \facnorm{y_{n,0}}{n}.
 \end{equation}
Now we collect all the estimates from (\ref{est1}) through (\ref{est8}) and obtain
\begin{eqnarray*}
\lefteqn{\facnorm{\perturb{z}_{n+2}}{n+1+\alpha} \leq \left( 1 + \frac{\facnorm{\theta'}{1}^2}{(n+\alpha)(n-1+\alpha)} \right)
\facnorm{\perturb{z}_n}{n-1+\alpha} +} %\label{relative bound} 
\\
&& +  \frac{\facnorm{\perturb\theta'}{\alpha}\Gamma(n)}{\Gamma(n+1+\alpha)} \left(2 \Gamma(\alpha) \facnorm{\theta'_0}{1} + (\Gamma(\alpha)
 + \Gamma(1+\alpha)) \facnorm{\theta'}{1}\right) \facnorm{z_{n,0}}{n} + %\label{absolute bound 1}
 \\
&& +  \frac{\tc^{1+\alpha}\Gamma(n) }{\Gamma(n+1+\alpha)} \left( \facnorm{\theta'}{1} \facnorm{y_n}{n} + \facnorm{\theta'_0}{1} \facnorm{y_{n,0}}{n}
\right). %\label{absolute bound 2}
\end{eqnarray*}
This shows $\perturb{z}_{n+2} \in F_{n+1+\alpha}$. The above
calculations and the bounds from Theorem \ref{upper bounds} now
imply the existence of a locally bounded function $\phi: \R^+ \to \R^+$ with $\phi(x) = \Or(x)$ as $x \to 0$, such that with $M = \max \{
\facnorm{\theta'}{1}, \facnorm{\perturb\theta'}{\alpha} \}$ and $Q =
\phi(M)$ we have
$$ \facnorm{\perturb{z}_{n+2}}{n+1+\alpha} \leq \left(1 + \frac{Q}{n(n-1)}\right) \facnorm{\perturb{z}_n}{n-1+\alpha} +
\frac{\Gamma(n)}{\Gamma(n+1+\alpha)} Q.$$
Moreover, since
\begin{equation} \label{gamma asymptotics}
\lim_{n \to \infty} n^{\beta} \frac{\Gamma(n)}{\Gamma(n+\beta)} = 1
\end{equation}
for each $\beta > 0$ and $n^{\beta}
\frac{\Gamma(n)}{\Gamma(n+\beta)} \leq 1$ for each $n \in \N$ and
$\beta \geq 1$ (cf. \cite{AbSt}, 6.1.46 and 6.1.47), the sequence
$(\facnorm{\perturb{z}_n}{n+\alpha-1})_{n \in \N}$ is bounded by the
sequence $(a_n)_{n \in \N}$ defined through
$$ a_{n+2} = a_n \left(1+\frac{2Q}{n^2}\right) + \frac{Q}{n^{1+\alpha}}$$
with $a_2 = \facnorm{\perturb
z_2}{1+\alpha} \leq \facnorm{\perturb\theta'}{\alpha}$. $a_n$ is
increasing, and so either $(Q+1) a_n  \leq Q$ for all $n$ (then $a_n
\leq Q$), or eventually $\frac{Q}{n^{1+\alpha}} \leq \frac{a_n
(Q+1)}{n^{1 + \alpha}}$, and then
$$ a_{n+2} \leq a_n \left(1 + \frac{2Q}{n^2} + \frac{Q+1}{n^{1+\alpha}} \right) \leq a_n \left(1 + \frac{3(Q+1)}{n^{1+\alpha}}\right).$$
This shows
$$ a_{n+2} \leq a_2 \prod_{k=1}^{n/2} \left( 1 + \frac{3(Q+1)}{(2k)^{1+\alpha}} \right) \leq \facnorm{\perturb\theta'}{\alpha}
\exp\left( 3(Q+1) \sum_{k=1}^\infty
\frac{1}{(2k)^{1+\alpha}}\right)$$
where the infinite sum is bounded uniformly in  $\alpha> \inf J_\alpha > 0$.
The last inequality above follows by taking the logarithm of the
 product above and using $\ln(1+|x|) < |x|$ in the resulting sum.
Thus we obtain
$$ \facnorm{\perturb{z}_{n+2}}{n+1+\alpha} \leq \max \left \{ \phi(M), \facnorm{\perturb\theta'}{\alpha}
\exp\left( 3(\phi(M)+1) \sum_{k=1}^\infty
\frac{1}{(2k)^{1+\alpha}}\right) \right\},$$
and the claim for $\perturb{z}_n$ is shown.

Turning now to the $\perturb{x}_n$, (\ref{xrec}) implies
\begin{equation} \label{xpert form}
\perturb{x}_n = -\I (\perturb z_{n-1}' - \theta'_0 \perturb y_{n-1} - \perturb \theta' y_{n-1}),
\end{equation}
and (\ref{diffeq}) gives
\begin{equation} \label{ypert form}
\perturb y_{n-1} = - \int_0^t (\perturb\theta' z_{n-1} + \theta'_0 \perturb z_{n-1}) \, ds - y_{n-1}(0) + y_{n-1,0}(0).
\end{equation}
The claim now follows in a very similar way as above from
Propositions \ref{key formulas 1} and \ref{Integration}. \qedi
\end{proof}

By the Lemma and Proposition \ref{key formulas 1} a),
$$ \norm[\infty]{\perturb z_n} \leq \facnorm{\perturb z_n}{n-1+\alpha} \frac{\Gamma(n+1-\alpha)}{\tc^{n+1-\alpha}}
 \leq \phi(M) \frac{(n-1)!}{\tc^n} \left( \frac{ \Gamma(n+1-\alpha)}{\tc^{1-\alpha} \Gamma(n)} \right).$$
Now (\ref{gamma asymptotics}) implies
$$ \norm[\infty]{\perturb z_n} \leq c \phi(M) \frac{(n-1)!}{\tc^n} n^{-1+\alpha},$$
with
$$c = \sup_{n \in \N, \alpha \in J_\alpha} \frac{ \Gamma(n+1-\alpha)}{\tc^{1-\alpha} \Gamma(n)} n^{1-\alpha} < \infty,$$
and (\ref{zpert}) is shown. The same reasoning applies to $\xi_n$,
showing (\ref{xpert}) and finishing the proof. \qedi
\end{proof}

\section{General Hamiltonians}

Our main result Theorem~\ref{MainTh} is formulated for Hamiltonians
\eqref{hamiltonian} with constant eigenvalues satisfying
Assumptions~1 or 2. In this section we show that these assumptions
are satisfied for a large class of Hamiltonians after transformation
to the natural time scale.

 Let us consider
\begin{equation} \label{tilde H 2}
\big( \I \eps \partial_s -\widetilde H(s)\big)  \psi(s) = 0
\end{equation}
for the traceless real-symmetric Hamiltonian
\begin{equation} \label{tilde hamiltonian 2}
\widetilde H(s) = \left( \begin{array}{cc}Z(s)&X(s)\\X(s)&-Z(s)
\end{array}
                                  \right)=\rho(s)\left( \begin{array}{cc} \cos\tilde \theta(s) &\,\,\, \sin\tilde\theta(s) \\
                                  \sin\tilde\theta(s) & \,-\hspace{-1pt} \cos\tilde\theta(s) \end{array}
                                  \right)\,.
\end{equation}
If $X^2+Z^2 > 0$, then for each $s_{\rm r}\in\R$ the transformation
\begin{equation}
 \tau(s) = 2 \int_{s_{\rm r}}^s \sqrt{\rho^2(u)} \, \D u \label{trafo2}
\end{equation}
takes the equation (\ref{tilde H 2}) with Hamiltonian (\ref{tilde
hamiltonian 2}) into equation (\ref{H}) with Hamiltonian
(\ref{hamiltonian}) with $\theta = \tilde\theta \circ \tau^{-1}$.
Berry and Lim \cite{BeLi} found that under very general conditions
on $X$ and $Z$  the singularities of $\theta'$ have the form of a
first order pole plus lower order singularities. Then, as to be
explained, by the Darboux' Principle Assumption~1 resp.\
Assumption~2 are satisfied pointwise on the real line. More
precisely, the $n^{\rm th}$ derivative of $\theta'$ at $t \in \R$
behaves like $\Gamma(n) r^{-n}$ as $n \to \infty$, where $r$ is the
distance from $t$ to the nearest pole; the corrections have
derivatives  going like $\Gamma(n-\alpha)r^{-n}$ as $n\to\infty$ for
some $\alpha
> 0$, and thus are in $F_{1-\alpha,r}$.

The task is now to make this discussion rigorous, and we start with
giving a version of  Darboux' Theorem. While this theorem in various
forms is certainly well known \cite{He,Bo,Di}, we were unable to
find a statement in the precision and generality we need in the
literature. The proof given in the Appendix uses Cauchy's formula
and explicit integration near the singularities.
 This strategy  was suggested to us by
Vassili Gelfreich.

\begin{theorem}[Darboux' Principle] \label{Darboux}
Let $f$ be analytic on $D_R=\{z\in\C:|z|<R\}$, and assume that $f$
is analytic also on $\partial D_R$ except for finitely many points.
Assume that there exists $N \in \N$ and $(z_j,\alpha_j,g_j)_{j \leq
N}$ with the properties:
\begin{itemize}
\item[(i)] For each $j$, $z_j$ is one of the  singularities of $f$ on $\partial D_R$;
\item[(ii)] $\alpha_j\in \R\setminus\{0,-1,-2,\ldots\}$;
\item[(iii)] the function $g_j$ is analytic in a neighborhood $U_j$ of $z=0$;
\item[(iv)] When $z_0$ is a singularity of $f$ on $\partial D_R$ and  $A_{z_0} := \{ j: z_j = z_0 \}$, then
\begin{equation} \label{singularities of f}
 f(z) = \sum_{j \in A_{z_0}}(z-z_0)^{-\alpha_j} \,g_j(z-z_0) \quad\mbox{on}\quad
\bigcap_{j \in A_{z_0}}(U_j+z_0) \cap D_R\, .
\end{equation}
\end{itemize}
Then
\begin{equation}\label{darbouxformel}
\frac{f^{(n)}(0)}{n!} =  \sum_{j=1}^N \E^{-\I\pi\alpha_j}
\frac{g_j(0)}{\Gamma(\alpha_j)}
\frac{n^{\alpha_j-1}}{z_j^{n+\alpha_j}}
\,\left(1+\Or\left({\textstyle\frac{1}{n}}\right)\right)\,.
\end{equation}
\end{theorem}

In words, Darboux' Theorem says that when $f$ has finitely many
algebraic convergence-limiting singularities, each of those
contributes to the growth of the derivatives of $f$ with a term of
absolute value $|g(0)| \Gamma(n+\alpha)/R^{n+\alpha}$ depending on
the strength $g(0)$, order $\alpha$ and distance $R$ of the
singularity, and a phase depending on its strength and location.
It is now clear that any function fulfilling the assumptions of Theorem
\ref{Darboux} is in $F_{\alpha,R}(\{0\})$, where $\alpha = \max_j
\alpha_j$. We have to prove this for the analytic continuation of
$\theta'$, and for this purpose put the following assumptions on the Hamiltonian
$\tilde H$:\\

\noindent {\bf Assumption XZ:} {\it $X$ and $Z$ are meromorphic in
an open set $U \subset \C$ containing some point $s_{\mr r}$ on the
real axis. $X$ and $Z$  fulfill
\begin{equation} \label{positive}
\rho^2(s) := X^2(s) + Z^2(s) \geq c > 0 \qquad \forall s \in \R\cap
U.
\end{equation}
By convention, we do not lift removable singularities of $\rho^2$,
so the critical points of $\rho^2$ consist of its zeros and the
poles of $X$ and $Z$. For $s$ close to such a critical point $s_0$
of $\rho^2$, we require
\begin{eqnarray}\label{formOfXZ}
X(s) = (s-s_0)^m f(s-s_0) \,\big[ 1 + (s-s_0)^{n} g_X(s-s_0)
\big], \nonumber\\ [-2mm]\\[-2mm]
Z(s) = \pm\I (s-s_0)^m f(s-s_0) \,\big[1 + (s-s_0)^{n} g_Z(s-s_0).
\big]\nonumber
\end{eqnarray}
where $0 < n \in \N$, $m \in \Z$ and $\frac{2m+n}{2}>-1$; the
functions $f$, $g_X$ and $g_Z$ are analytic in a neighborhood of
$s_0$, and $K := |f^2(0)(g_X(0)-g_Z(0))| > 0$.
The set of critical points has no accumulation points in $U$.}\\

The class of $X$ and $Z$ fulfilling Assumption XZ is  smaller than
the universality class considered in \cite{BeLi}. It is not our
ambition here to investigate just how large we can take our class
while still giving a mathematically rigorous proof; but note that
(\ref{formOfXZ}) does contain the generic case of analytic $X$ and
$Z$ and a simple zero of $\rho^2$, i.e.\ $m=0$ and $n=1$, along with
many others.

By (\ref{positive}), the map $s \mapsto \sqrt{\rho(s)^2}$ is
analytic and invertible on $U \cap \R$, but may have critical points
in the complex plane. For each such critical point $s_0$ for which a
branch cut $B_{s_0}$ of $\sqrt{\rho^2}$ is needed, we choose the
branch cut such that it points away from $s_{\mr r}$, and define
$U_0$ to be the connected component of $U \setminus \bigcup_{s_0}
B_{s_0}$ containing $s_{\mr r}$. Then $\tau$ as given in
(\ref{trafo2}) is well-defined and analytic in $U_0$. We define
$$ C_r(s_{\mr r}) := \{ s \in U_0: |\tau(s)-\tau(s_{\mr r})| < r \},$$
denote by $C_{r,0}$ the connected component of $C_r$ containing
$s_{\mr r}$ and put
$$R = R(s_{\mr r}) := \sup \{r > 0: \overline{C_{r,0}(s_{\mr r})} \subset U_0 \}.$$
Figure 1 shows some typical cases.

\begin{figure}%[htp]
\centering %\epsfig{file=beulen.ps, width=0.7\textheight, scale=1}
\includegraphics[width=\textwidth]{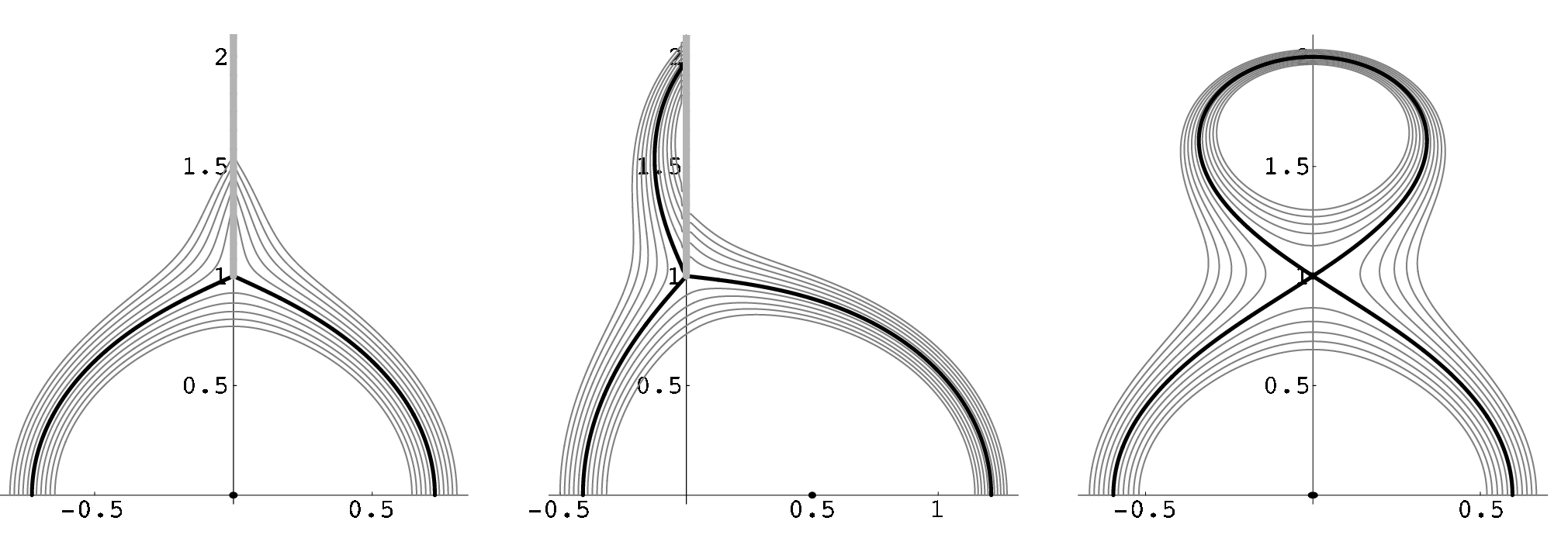}
\caption{\em The boundaries of the sets $C_r$ for various
situations; the black line is the boundary of $C_R$. In (a) and (b),
$\rho^2 = 1+s^2$, and $s_{\mr r}=0$ in (a) while $s_{\mr r}=0.5$ in
(b). The branch cuts are on the imaginary axis. In (c), $\rho^2 =
(1+s^2)^2$ and $s_{\mr r}=0$;  examples (b) and (c)  show why it is
necessary to consider the connected components.} \vspace*{-7.5cm}
\leftline{ (a) \hspace{0.3 \textwidth } (b) \hspace{0.3 \textwidth} (c)}
\vspace*{6.5cm}
\end{figure}

We restrict the discussion to the case where $C_r(s_{\mr r})$ hits
the boundary of $U_0$ at  critical points of $\rho^2$ as $r$ grows
to $R$. When $U$ is large enough, this is the generic case, provided
one has put the branch points $B_{s_0}$ in a
sensible way. We distinguish two cases corresponding to our Assumptions 1 and 2 on $\theta'$:\\

\noindent {\bf Assumption R1:} {\it Let $s_{\mr r}\in \R$ be a point
where a Stokes line of $\tau$, i.e.\ a level line of Re$(\tau(s))$,
emanating from a critical point $s_0$ of $\rho^2$ crosses the real
axis. Assume that $\partial C_R(s_{\mr r}) \cap \partial U_0 =
\{s_0, \overline{s_0} \}$, which means that $\tau(s_0)$ is closer to
$\tau(s_{\rm r})$ than the
$\tau$-image   of any other critical point   of $\rho^2$.}\\

\noindent {\bf Assumption R2:} {\it $\partial C_R(s_{\mr r}) \cap \partial U_0 = \{s_1, \ldots s_k\}$, where the $s_j$ are critical points of $\rho^2$.}\\

\begin{theorem}
Let $\tilde H$, $\rho$ and $\tilde \theta$ be defined as in
(\ref{tilde H 2}) with $X$ and $Z$ fulfilling Assumption~XZ. Define
$\theta = \tilde \theta \circ \tau^{-1}$ with $\tau$  given by
(\ref{trafo2}), and $t_0 = \tau(s_0)$.
\begin{enumerate}\rom
\item If Assumption R2 holds, then $\theta' \in F_{1,R}(\{\tr\})$, i.e.\ Assumption 2 is fulfilled at $\tr$.
\item Suppose that Assumption R1 holds, and that $X$ and $Y$ are given by (\ref{formOfXZ}) at $s_0$.
Then with $\gamma = \frac{\pm n }{2m+n+2}$ there exists a closed
interval $I \ni s_{\mr r}$ such that on $I$,
$$ \theta'(t) = \I\gamma\left( \frac{1}{t-t_0} - \frac{1}{t-\overline{t_0}}\right) + \perturb\theta'(t) \,.$$
where $\perturb \theta' \in F_{\alpha,R}(I)$ for each $\alpha \in
(0,1)$ with $\alpha \geq \frac{2m + n}{2m + n + 2}$.
\end{enumerate}
\end{theorem}

\begin{proof}
Let Assumption 2 hold; without loss we assume $s_{\mr r} = 0$. Then
$\tau$ is analytic on $C_{R}$, while on $\partial C_{R}$ there are
finitely many singularities. Let $s_0$ be a singularity, and let
$A_0$ be the class of functions $h$ which are analytic in a
neighborhood of $0$ with $h(0) = 0$. Then for $s$ close to $s_0$,
$$\rho^2(s) =  2 K (s-s_0)^{2m+n} (1 + h_1(s-s_0))$$
with $h_1 \in A_0$. Consequently
\begin{eqnarray} \label{t_sigma}
\tau(s)-\tau(s_0) &=& 2 \int_{s_0}^s (r-s_0)^{(2m+n)/2} \sqrt{2K(1 + h_1(r-s_0))} \, dr = \nonumber \\
%&=&  2 \int_{0}^{s-s_0} r^{(2m+n)/2} \sqrt{2K(1 + h_1(r))} \, dr = \\
&=&\frac{4\sqrt{2K}}{2m+n+2} (s-s_0)^\frac{2m+n+2}{2}
(1+h_2(s-s_0))
\end{eqnarray}
with $h_2 \in A_0$.

Since by construction $\tau$ has no critical points inside $C_R$, it
is locally analytically invertible there. Since $D_R:=\tau(C_R)$ is
the disc with radius $R$ and center $\tau(s_{\rm r})$, global
invertibility follows. Thus $\tau$ is one-to-one from $C_{R}$ onto
$D_{R}$ with analytic inverse.

We have
$$\theta'(\tau(s)) = \frac{\tilde\theta'(s)}{2 \rho(s)} = \frac{1}{2 \rho(s)} \frac{\D}{\D s} \arctan
 \left(\frac{X}{Z}\right)(s) = \frac{X'Z - Z'X}{2 \rho^3}(s),$$
and taking $s = \tau^{-1}(t)$, $ t \in D_R$, shows that $\theta'$ is
analytic on the circle $D_R$, fulfilling the first assumption of Darboux'
theorem. Note in particular that $X'Z - Z'X$ is non-singular on
$D_R$ by our convention of not lifting removable singularities of
$\rho^2$. For the behavior of $\theta'$ near a singularity $s_0$ at the boundary of $D_R$, we revisit the calculation of Berry and Lim \cite{BeLi}, paying special attention to the error terms that arise.
We have
$$
(X'Z-Z'X)(s) = \pm \I n K (s-s_0)^{2m+n-1} \big( 1 + h_3(s-s_0) \big),
$$
with $h_3 \in A_0$, $A_0$ as above. Writing $\sigma = s-s_0$ we now obtain
\begin{eqnarray*}
\theta'(\tau(s)) &=& \frac{\pm \I n K  \sigma^{2m+n-1} ( 1 + h_3(\sigma) )}{2 (2K \sigma^{2m+n})^{3/2}
(1+h_1(\sigma))^{3/2}} = \frac{\pm \I n ( 1 + h_3(\sigma) )}{4 \sqrt{2K} \sigma^{(2m+n+2)/2} (1+h_1(\sigma))^{3/2}}\\
&=& \frac{\pm \I n }{(2m+n+2)(\tau(s)-\tau(s_0))} \frac{( 1 + h_3(\sigma))(1+h_2(\sigma))}{ (1+h_1(\sigma))^{3/2} } \\
&=& \frac{\pm \I n }{(2m+n+2)(\tau(s)-\tau(s_0))} (1 + h_4(\sigma)),
\end{eqnarray*}
with $h_4 \in A_0$, where we used (\ref{t_sigma}) in the second
line. It remains to find the form of $\tau^{-1}$ near the
singularity $\tau(s_0)$. First note that since $\tau$ is an
integral, $\tau(s)-\tau(s_0)=\tau(\sigma) $. From (\ref{t_sigma}),
$\tau(\sigma) = \tilde K \sigma^{\alpha}(1+h_2(\sigma))$ with the
obvious $\tilde K$ and $\alpha$. The function
$$\sigma \mapsto \sigma \tilde K^{1/\alpha} (1+h_2(\sigma))^{1/\alpha}$$ is invertible in a neighborhood
of $\sigma = 0$, and the inverse function $h_5(\sigma)$ is an element of $A_0$.
%Since
%$$ h_5(t(\sigma)^{1/\alpha}) = h_5(\tilde K^{1/\alpha} (1+h_2(\sigma))^{1/\alpha}) = \sigma,$$
We then find
$$ h_4(\sigma) = h_4 \circ h_5 \left(\tau(\sigma)^{2/(2m+n+2)}\right) = \sum_{j=1}^{2m+n+2} \tau(\sigma)^{\frac{2j}{2m+n+2}} g_j(\tau(\sigma))$$
with analytic functions $g_j$.
Putting things together and writing $t = \tau(s)$, $t_0 = \tau(s_0)$ we obtain
\begin{equation} \label{thetastrich}
 \theta'(t) = \frac{\pm \I n }{(2m+n+2)(t-t_0)} \left(1 + \sum_{j=1}^{2m+n+2} (t-t_0)^{\frac{2j}{2m+n+2}} g_j(t-t_0)\right)
 \end{equation}
in a neighborhood of $t_0$. This has exactly the form
(\ref{singularities of f}),  and thus Darboux' Theorem shows (i). As
for (ii), note that when Assumption R1 is fulfilled, by continuity
of $\tau$ still $\partial C_R(s) \cap \partial U_0 = \{s_0,
\overline{s_0} \}$  for all $s$ in a real neighborhood of $s_{\mr
r}$. All the above calculations only need information from the
singularities, so they are valid without change, and the proof is
finished. \qedi
\end{proof}

\begin{example} {\bf (Landau Zener transitions):}
In the Landau-Zener model, $X(s) = s, Z(s) = \delta$ and
consequently $\rho(s) = \sqrt{\delta^2+s^2}$. The critical points of
$\rho$ are at $\pm \I \delta$, and
$$R(0) = \tau(\I \delta) = 2 \int_0^\delta \sqrt{\delta^2-s^2} \, \D s = \frac{\pi \delta^{3/2}}{2}.$$
Moreover, at $s = \pm \I \delta$ the functions $X$ and $Z$ have the
form (\ref{formOfXZ}) with $m=0$, $n=1$, $f = \pm \I \delta$, $g_X =
\mp \I / \delta$ and $g_Z = 0$. Thus
$$ \theta'(t) = \pm \frac{\I}{3 (t \mp \I \delta)}\left(1 + (t \mp \I\delta)^{2/3} g_1(t \mp \I\delta) +  (t \mp \I\delta)^{4/3} g_2(t \mp \I\delta)\right),$$
where $g_1$ and $g_2$ are analytic near $\pm \I \delta$. Thus
$$\theta'_0 := \frac{\I}{3 t-\I\delta} - \frac{\I}{3 t+\I\delta}$$ and $\perturb\theta = \theta' - \theta_0'  \in
F_{1/3,\delta}(I)$ for some $I \supset \{ 0 \}$. In the simple situation at hand,
it is easy to see  that in fact
$\perturb\theta' \in F_{1/3,\delta}(I)$ for every finite interval
$I$. When $\rho^2$ has more that one critical point on each side of
the real axis, the situation is more involved and we refer to
\cite{BeTe2}.
\end{example}

\section*{Appendix}

\subsection*{Proof of Lemma \ref{CD bound} }
%\begin{Proof}[Proof of Lemma \ref{CD bound}]
We start by converting (\ref{CC}) and (\ref{DD}) to a recursion
for the $D_n$ alone by plugging (\ref{C from D}) into (\ref{DD}). The result, in a somewhat expanded form, is
\begin{eqnarray}
D_n &=& \frac{\theconstant^2}{4} \sum_{k=1}^{n-1} B(k,n-k) + \label{z1}\\
&& +\, \frac{\theconstant^2}{2} \sum_{k=2}^{n-1} B(k,n-k) \left( \sum_{j=1}^{k-1} \frac{D_j}{j} + \sum_{j=1}^{n-k-1} \frac{D_j}{j} \right) + \label{z2}\\
&& + \,\theconstant^2 \sum_{k=2}^{n-1} B(k,n-k) \sum_{j=1}^{k-1} \frac{D_j}{j} \sum_{l=1}^{n-k-1}\frac{D_l}{l} + \label{z3} \\
&& + \,\sum_{k=2}^{n-2} B(k,n-k) D_k D_{n-k}. \label{z4}
\end{eqnarray}
To show (\ref{D estimate}), we will of course proceed inductively.
Direct calculation yields that (\ref{D estimate}) is true up to
$n=10$ (even for $M = 1$). Let us now suppose that $n \in \N$ is an
even number and that (\ref{D estimate}) holds up to $n-2$. We will
show that (\ref{D estimate}) also holds for  $n$ and for this
purpose treat each line of (\ref{z1}) through (\ref{z4}) separately.
We start with (\ref{z4}). Using the induction hypothesis, we get
\begin{eqnarray*}\lefteqn{
(\ref{z4}) \leq  \sum_{k=2}^{n-2} B(k,n-k) \left( \tfrac{1}{k-1} 
\sum_{j=1}^{\floor{k/2}}\tfrac{\theconstant^{2j}M^j}{j!}\right)
 \left( \tfrac{1}{n-k-1} \sum_{l=1}^{\floor{(n-k)/2}}  \tfrac{\theconstant^{2l}M^l}{l!} \right) =} \\
&=& \tfrac{1}{(n-1)(n-2)}\sum_{k=2}^{n-2} B(k-1,n-k-1)  \sum_{j=1}^{\floor{k/2}}\tfrac{\theconstant^{2j}M^j}{j!}
 \sum_{l=1}^{\floor{(n-k)/2}}\tfrac{\theconstant^{2l}M^l}{l!} = (\ast_1).
\end{eqnarray*}
We sort this triple sum by powers $p$ of $\theconstant^2$, i.e. take $p = j+l$. The scheme is the following:
$$ \begin{array}{lllll}
p=2: & j=1 & \Rightarrow & l=1, & k = 2,3, \ldots, n-2. \\
p=3:  & j=1 & \Rightarrow & l=2, & k = 2,3, \ldots, n-4 \\
     & j=2 & \Rightarrow & l=1, & k = 4,5, \ldots, n-2\\
     \vdots
\end{array}
$$
so for given $p \leq n/2$ (which is the highest power of
$\theconstant^2$ that occurs) we have $j$ running form $1$ through
$p-1$, and for this $j$ we have $l=p-j$ and $k=2j, 2j+1, \ldots,
n-2(p-j)$. This gives
\begin{eqnarray*}
(\ast_1) &=& \tfrac{1}{(n-1)(n-2)}
 \sum_{p=2}^{n/2} \theconstant^{2p}M^p \left( \sum_{j=1}^{p-1} \tfrac{1}{j!(p-j)!} \sum_{k=2j}^{n-2p+2j} B(k-1,n-k-1) \right) = \\
&=& \tfrac{1}{(n-1)(n-2)} \sum_{p=2}^{n/2}
\tfrac{\theconstant^{2p}M^p}{p!}  \underbrace{\left( \sum_{j=1}^{p-1}
\sum_{k=0}^{n-2p} \binkoeff{p}{j}
\frac{B(2j+k-1,n-2j-k-1)}{j!(p-j)!} \right)}_{\psi(n,p)}.
\end{eqnarray*}
We will prove that $\psi(n,p)$ is bounded uniformly in $n$ and $p \leq n/2$. We use the integral representation
\begin{equation} \label{intrep}
B(n,m) = \int_0^1 u^{m-1}(1-u)^{n-1} \, \D u
\end{equation}
of the Beta function, and obtain
\begin{eqnarray*}
\psi(n,p) &=& \int_0^1 \sum_{j=1}^{p-1} \sum_{k=0}^{n-2p} \binkoeff{p}{j}u^{2j+k-2}(1-u)^{n-2j-k-2} \, \D u = \\
&=& \int_0^1 \left( \sum_{j=1}^{p-1} \binkoeff{p}{j} u^{2j-2}(1-u)^{2(p-j)-2} \right) \left( \sum_{k=0}^{n-2p} u^k (1-u)^{n-2p-k} \right) \,\D u.
\end{eqnarray*}
The sum in the second bracket above is bounded by $2$ uniformly on $[0,1]$, and thus $\psi(n,p)$ is shown to be bounded provided we are able to prove
\begin{equation} \label{binkoeff formula}
\int_0^1 \sum_{j=1}^{p-1} \binkoeff{p}{j} u^{2j-2}(1-u)^{2(p-j)-2} \D u \leq 5.
\end{equation}
To see (\ref{binkoeff formula}), first note that the terms with
$j=1$ and $j=p-1$ are equal to $p/(2p-3)$ and thus bounded by $2$
since $p \geq 2$. For the remaining terms, we use $u^{2j-2} \leq
u^j$ and $(1-u)^{2(p-j)-2} \leq (1-u)^{p-j}$.We then obtain
$$\sum_{j=2}^{p-2} \binkoeff{p}{j} u^{2j-2}(1-u)^{2(p-j)-2} \leq  \sum_{j=0}^{p} \binkoeff{p}{j} u^{j}(1-u)^{p-j} = 1$$
by the binomial theorem. This proves (\ref{binkoeff formula}), and we obtain
\begin{equation} \label{z4 estimate}
(\ref{z4}) \leq \frac{5}{(n-1)(n-2)} \sum_{p=2}^{n/2} \frac{\theconstant^{2p}M^p}{p!} \leq \frac{1}{2(n-1)} \sum_{p=2}^{n/2} \frac{\theconstant^{2p}M^p}{p!}
\end{equation}
since $n \geq 12$.
We now turn to (\ref{z3}), and start by proving
\begin{lemma}
For $m < n$ we have
\begin{equation} \label{step1}
\sum_{j=1}^{m}\frac{D_j}{j} \leq \frac{1}{m} \sum_{l=1}^{\floor{m/2}} \frac{(m-2l+1)\theconstant^{2l}M^l}{(2l-1)l!}.
\end{equation}
\end{lemma}
\begin{proof}
We use the induction hypothesis to calculate
$$\sum_{j=1}^m \frac{D_j}{j}  \leq  \sum_{j=2}^m \frac{1}{j(j-1)} \sum_{l=1}^{\floor{j/2}}\frac{\theconstant^{2l}M^l}{l!}
 =  \sum_{l=1}^{\floor{m/2}}\frac{\theconstant^{2l}M^l}{l!} \sum_{j=2l}^{m} \frac{1}{j(j-1)}. $$
The claim now follows from
\begin{equation} \label{easysum}
\sum_{k=j}^{m} \frac{1}{k(k-1)} = \frac{1}{j-1} - \frac{1}{m} = \frac{m-j+1}{m(j-1)}.
\end{equation}
\end{proof} \qedi

Using (\ref{step1}) we get
\begin{eqnarray*}
 (\ref{z3}) &\leq& \theconstant^2 \sum_{k=2}^{n-1} \frac{B(k,n-k)}{(k-1)(n-k-1)}
 \sum_{j=1}^{\floor{\frac{k-1}{2}}}\frac{(k-2j)\theconstant^{2j}M^j}{(2j-1) j!} \times \\
 && \times
\sum_{l=1}^{\floor{\frac{n-k-1}{2}}}\frac{(n-k-2l)\theconstant^{2l}M^l}{(2l-1)l!} =: (\ast_2).
\end{eqnarray*}
Again we sort this by powers $p$ of $\theconstant^2$, leaving the leading $\theconstant^2$ out.
The scheme is
$$ \begin{array}{lllll}
p=2: & j=1 & \Rightarrow & l=1, & k = 3,4, \ldots, n-3. \\
p=3  & j=1 & \Rightarrow & l=2, & k = 3,4, \ldots, n-5 \\
     & j=2 & \Rightarrow & l=1, & k = 5,6, \ldots, n-3,\\
     \vdots
\end{array}
$$
and this time the general term is $j = 1, \ldots, p-1$, $l=p-j$, $k=2j+1, \ldots, n-2(p-j)-1.$
We use $\frac{B(k,n-k)}{(k-1)(n-k-1)} = \frac{B(k-1,n-k-1)}{(n-1)(n-2)}$ and obtain
\begin{equation} \label{phieq}
(\ast_2) =  \frac{\theconstant^2}{(n-1)(n-2)} \sum_{p=2}^{n/2-1} \frac{\theconstant^{2p} M^p}{p!} \phi(n,p)
\end{equation}
with
\begin{eqnarray*}
\lefteqn{\phi(n,p) = \sum_{j=1}^{p-1} \sum_{k=2j+1}^{n-2p+2j-1} \binkoeff{p}{j}\frac{(k-2j)(n-k-2(p-j))}{(2j-1)(2(p-j)-1)}  B(k-1,n-k-1) =} \\
&=& \sum_{j=1}^{p-1} \sum_{k=1}^{n-2p-1} \binkoeff{p}{j}\frac{k(n-k-2p)}{(2j-1)(2(p-j)-1)}  B(2j + k-1,n-2j -k-1) = \\
&=& \int_0^1 \sum_{j=1}^p \binkoeff{p}{j} \tfrac{u^{2j-1}(1-u)^{2(p-j)-1}}{(2j-1)(2(p-j)-1)} \sum_{k=1}^{n-2p-1}k(n-2p-k)u^{k-1}(1-u)^{n-2p-k-1} \, \D u.
\end{eqnarray*}
The second sum above is obviously bounded by $(n-2p-1)
\sum_{k=1}^{\infty} k u^{k-1} \leq 4(n-2p-1)$ on $[0,1/2]$, and
since it is symmetric around $u=1/2$, this bound is also valid on
$[0,1]$. On the other hand, the integral representation of the Beta
function yields
\begin{eqnarray*}
&&\int_0^1 \sum_{j=1}^p \binkoeff{p}{j} \frac{u^{2j-1}(1-u)^{2(p-j)-1}}{(2j-1)(2(p-j)-1)} \, \D u =
\sum_{j=1}^{p-1} \binkoeff{p}{j} \frac{B(2j,2(p-j)}{(2j-1)(2(p-j)-1)} =\\
&&\hspace{1cm}= \frac{1}{(2p-1)(2p-2)} \sum_{j=1}^{p-1} \binkoeff{p}{j} B(2j-1,2(p-j-1)) =\\ 
&&\hspace{1cm}= \frac{1}{(2p-1)(2p-2)}
 \int_0^1 \sum_{j=1}^{p-1}\binkoeff{p}{j} u^{2j-2}(1-u)^{2(p-j)-2} \, \D u.
\end{eqnarray*}
The last integral is bounded by $5$ due to (\ref{binkoeff formula}), and thus
$\phi(n,p) \leq \frac{20(n-2p-1)}{(2p-1)(2p-2)} \leq \frac{5(n-2)}{(p+1)}$. Inserting in (\ref{phieq}) yields
\begin{equation} \label{z3 estimate}
(\ref{z3}) \leq \frac{5}{M(n-1)} \sum_{p=2}^{n/2-1} \frac{\theconstant^{2(p+1)} M^{p+1}}{(p+1)!}.
\end{equation}
Turning to (\ref{z2}), note first that for $j < n-1$, we have
\begin{eqnarray*}
 \sum_{k=j+1}^{n-1} B(k,n-k) &=& B(j+1,n-j-1) + \\
 &&+B(j+2,n-j-2) + \ldots + B(n-1,1) =\\
 &=& \sum_{k=1}^{n-j-1} B(k,n-k).
 \end{eqnarray*}
Moreover, $B(1,n-1) = 1/(n-1)$ and $B(k,n-k) \leq 2/((n-1)(n-2))$ for $2 \leq k \leq n-2$, and thus
$$ \sum_{k=1}^{n-1}B(k,n-k) \leq 4/(n-1).$$
By symmetry,
\begin{eqnarray*}
(\ref{z2}) &=& 2 \theconstant^2 \sum_{k=2}^{n-1} B(k,n-k) \sum_{j=1}^{k-1} \frac{D_j}{j} 
= 2 \theconstant^2 \sum_{j=1}^{n-2} \frac{D_j}{j} \sum_{k=j+1}^{n-1} B(k,n-k) \leq\\
&\leq& \frac{8\theconstant^2}{n-1} \sum_{j=1}^{n-2} \frac{D_j}{j} \leq \frac{8\theconstant^2}{(n-1)(n-2)}
\sum_{l=1}^{n/2-1} \frac{(n-2l)\theconstant^{2l}M^l}{(2l-1)l!}
\end{eqnarray*}
where the last inequality is (\ref{step1}). Thus
\begin{equation} \label{z2 estimate}
(\ref{z2}) \leq \frac{16}{M(n-1)}\sum_{p=2}^{n/2} \frac{\theconstant^{2p}M^p}{p!}.
\end{equation}
Finally
\begin{equation} \label{z1 estimate}
(\ref{z1}) = \frac{1}{4} \theconstant^2 \sum_{k=1}^{n-1} B(k,n-k) \leq \frac{\theconstant^2}{n-1}.
\end{equation}
Combining (\ref{z4 estimate}), (\ref{z3 estimate}), (\ref{z2 estimate}) and (\ref{z1 estimate}) we arrive at
$$ D_n \leq \frac{1}{n-1} \left( \theconstant^2 + \left(\frac{1}{2} + \frac{5}{M}\right) \frac{\theconstant^4M^2}{2!} +
 \sum_{p=3}^{n/2} \left(\frac{1}{2} + \frac{21}{M}\right) \frac{\theconstant^{2p}M^p}{p!} \right).$$
Choosing $M \geq 42$, the proof of Lemma \ref{CD bound} is finished.
%\end{proof}

\subsection*{Proof of Theorem \ref{Darboux}}

Let $z_0$ be a singularity of $f$ and write $U_{z_0} = \bigcap_{j\in A_{z_0}}(U_j)$. From (\ref{singularities of f}) it is clear that the function $z \mapsto \sum_{j \in A_{z_0}}
(z-z_0)^{-\alpha_j} \,g_j(z-z_0)$ is the unique analytic
continuation of $f$ from $(U_{z_0}+z_0) \cap D_R$ to $(U_{z_0}+z_0)
\setminus B_{z_0}$, where $B_{z_0} :=\{z\in\C: z= a z_0,\,a>1\}$ is
the branch cut (if necessary). Moreover $f$ is analytic on the
closed set $\overline{D_R} \setminus \bigcup_j (U_j+z_j)$, and therefore analytic in a neighborhood of that set. Putting this continuation together with the continuations near each singularity, we conclude that there exists $\delta > 0$ such that the analytic
continuation of $f$ to $D_{R+ 2 \delta} \setminus \bigcup_j B_j$
exists and is bounded on $(\partial D_{R+  \delta}) \setminus
\bigcup_j B_j$. Here we may choose $\delta < 1$ and sufficiently small to guarantee
 $D_{\delta} \in U_j$ and $D_{\delta R} \in U_j$  for all $j$.
Let $\Gamma = \partial D_{R+\delta} \cup \bigcup_{j=1}^N C_j$ be the
piecewise smooth path that encircles $0$ anticlockwise  along the
boundary of the disk with radius $R+\delta$ and avoids the branch
cuts $B_{z_0}$ by encircling clockwise the singularities at $z_j$ with a
circle $C_j$ of radius $\delta$.  Then
\[
\frac{f^{(n)}(0)}{n!} = \frac{1}{2\pi\I}\oint\limits_\Gamma
\frac{f(z)}{z^{n+1}}\,\D z = \frac{1}{2\pi\I}\int\limits_{\partial
D_{R+\delta}} \frac{f(z)}{z^{n+1}}\,\D z +
\frac{1}{2\pi\I}\,\sum_{j=1}^N\,\int\limits_{C_j}
\frac{f(z)}{z^{n+1}}\,\D z\,.
\]
The first integral is easily estimated through
\[
\left|\frac{1}{2\pi\I}\int_{\partial D_{R+\delta}}
\frac{f(z)}{z^{n+1}}\,\D z\right|\leq
\frac{\sup_{|z|=R+\delta}|f(z)|}{(R+\delta)^n} = R^{-n}
\Or(n^{-k})\quad\forall\,k\in\N\,.
\]
For the contribution of the poles, first note that by
(\ref{singularities of f}) we may treat each $j \leq N$ separately,
even if they belong to the same pole. If $\alpha_j\in \N$, a
straightforward computation shows that
\[
{\rm Res}_{z=z_j}\left( \frac{f(z)}{z^{n+1}}\right) =
(-1)^{\alpha_j-1} \frac{g_j(0)}{\Gamma(\alpha_j)}
\frac{n^{\alpha_j-1}}{z_j^{n+\alpha_j}}
\,\left(1+\Or\left({\textstyle\frac{1}{n}}\right)\right)\,.
\]
Noting that $C_j$ is negatively oriented,  we conclude that the
contribution from poles is the one claimed in \eqref{darbouxformel}.

For the remaining terms let $g_j(z-z_j) = \sum_{k=0}^\infty b_k
(z-z_j)^k$. Then
\begin{eqnarray}
\frac{1}{2\pi\I}\,\int\limits_{C_j} \frac{f(z)}{z^{n+1}}\,\D z
&=&\frac{1}{2\pi\I}\sum_{k=0}^\infty\,b_k \int\limits_{C_j}
\frac{(z-z_j)^{k-\alpha_j}}{z^{n+1}}\,\D z = \nonumber\\
&=&\frac{1}{2\pi\I}\sum_{k=0}^\infty\,b_k
z_j^{k-\alpha_j-n}\int\limits_{C_j/z_j}
\frac{(\zeta-1)^{k-\alpha_j}}{\zeta^{n+1}}\,\D \zeta  \,,\label{singleterm}
\end{eqnarray}
where we substituted $z=z_j\,\zeta$. The remaining integral is
shifted to the origin and can then be solved explicitly in terms of
the hypergeometric function $F$,
\begin{eqnarray*}
\int\limits_{C_j/z_j} \frac{(\zeta-1)^{k-\alpha_j}}{\zeta^{n+1}}\,\D
\zeta&=& \int\limits_{C_j/z_j-1}
\frac{\zeta^{k-\alpha_j}}{(\zeta+1)^{n+1}}\,\D \zeta \\&=&
\frac{F\left( \begin{array}{c} 1+n, k-\alpha_j+1\\
k-\alpha_j+2\end{array}; -\delta\right)}{k-\alpha_j+1} \left[
z^{k-\alpha_j+1} \right]^{\delta+0\I}_{\delta-0\I}\\&=& \frac{F\left(  \begin{array}{c}  1+n, k-\alpha_j+1\\
k-\alpha_j+2\end{array}; -\delta\right)}{k-\alpha_j+1}
\delta^{k-\alpha_j+1}\left( 1-\E^{-2\pi\I\alpha_j}\right)\,.
\end{eqnarray*}
In the second line above we used the power series expansion $$\frac{1}{(1+\zeta)^{n+1}} = \sum_{j=0}^{\infty} (-\zeta)^j \binkoeff{n+j}{j},$$ valid for $|\zeta| = \delta < 1$.  
Note also that the branch cut was moved to the positive real axis through the two changes of variables.

For $k\leq \max(a,2-a)$ we use the asymptotic expansion of the
hypergeometric function (cf.\ \cite{AbSt}, 15.7.2 )
\begin{eqnarray*}
 F\left( \begin{array}{c} 1+n, k-\alpha_j+1\\
k-\alpha_j+2\end{array}; -\delta\right)  = (\delta
n)^{-(k-\alpha_j+1)}
\Gamma(k-\alpha_j+2)\left(1+\Or\left(\textstyle\frac{1}{n}\right)\right)\,,
\end{eqnarray*}
which shows that from the first $\lfloor\max(a,2-a)\rfloor+1$ terms
in each sum only the $k=0$ terms contribute to the leading order in
\eqref{darbouxformel} with
\[
 b_0 \frac{n^{\alpha_j-1}}{z_j^{n+\alpha_j}}
\frac{\Gamma(2-\alpha_j)}{1-\alpha_j} \frac{ \E^{-\I\pi \alpha_j
}\sin((1-\alpha_j)\pi)}{\pi}= g_j(0)
\frac{n^{\alpha_j-1}}{z_j^{n+\alpha_j}}\frac{\E^{-\I\pi\alpha_j}}{\Gamma(\alpha_j)}\,.
\]
As to be shown, for $k>\max(a,2-a)$ we have
\begin{equation}\label{F estimate}
 F\left( \begin{array}{c} 1+n, k-\alpha_j+1\\
k-\alpha_j+2\end{array}; -\delta\right)  =
\Or\left(\frac{k-\alpha_j+1}{n^{1+\frac{k-\alpha_j}{2}}}\right) =
\Or\left(\frac{k-\alpha_j+1}{n^{2-\alpha_j}}\right)\,.
\end{equation}
Hence, these terms do no contribute to the leading order in
\eqref{darbouxformel}. Note that the sum over $k$ in
\eqref{singleterm} converges since $D_{\delta |z_j|} \in U_j$.

To check \eqref{F estimate} we use the integral representation of
the hypergeometric function,
\begin{eqnarray*} \lefteqn{
F\left( \begin{array}{c} 1+n, k-\alpha_j+1\\
k-\alpha_j+2\end{array}; -\delta\right) =
\frac{\Gamma(2-\alpha_j+k)}{\Gamma(1-\alpha_j+k)\Gamma(1)}
\int_0^1\D t \,\frac{t^{k-\alpha_j}}{(1+\delta t)^{n+1}}}\\
&=& (k-\alpha_j+1)\left( \int_0^s \D
t\,\frac{t^{k-\alpha_j}}{(1+\delta t)^{n+1}} +\int_s^1 \D
t\,\frac{t^{k-\alpha_j}}{(1+\delta t)^{n+1}}\right)\\&\leq&
(k-\alpha_j+1)\left(\left.
-\frac{s^{k-\alpha_j}}{n\delta}\frac{1}{(1+\delta t)^n}\right|_0^s +
\frac{1}{(1+\delta s)^{n+1}} \frac{1}{k-\alpha_j+1}
\left.t^{k-\alpha_j+1}\right|_s^1\right)\\ &\leq&
(k-\alpha_j+1)\frac{s^{k-\alpha_j}}{n\delta} +\frac{1}{(1+\delta
s)^{n+1}}   \\
&\stackrel{s = \frac{1}{\sqrt{n}}}{=}&
\frac{(k-\alpha_j+1)}{\delta}\frac{1}{n^{1+\frac{k-\alpha_j}{2}}} +
\frac{1}{(1+\frac{\delta
\sqrt{n}}{n})^{n+1}}=\Or\left(\frac{k-\alpha_j+1}{n^{1+\frac{k-\alpha_j}{2}}}\right)\,.
\end{eqnarray*}
%The proof is finished.


\begin{thebibliography}{99}

\bibitem[AbSt]{AbSt} M.\ Abramowitz and I.A.\ Stegun (Eds.). {\em Handbook of Mathematical Functions}, $9^{\rm th}$ printing, Dover, New York,
1972.

\bibitem[Ber]{Be}  M.\ V.\ Berry. {\em Histories of adiabatic quantum transitions},
Proc.\ R.\ Soc.\ Lond.\ A {\bf 429}, 61--72 (1990).

\bibitem[BerLi]{BeLi}  M.\ V.\ Berry and R.\ Lim. {\em Universal transition prefactors
derived by superadiabatic renormalization}, J.\ Phys.\ A {\bf 26},
4737--4747 (1993).

\bibitem[BeTe$_1$]{BeTe1} V.\ Betz and S.\ Teufel. {\em Precise coupling terms in adiabatic quantum evolution}, to appear in Annales Henri Poincar\'e (2004).

\bibitem[BeTe$_2$]{BeTe2} V.\ Betz and S.\ Teufel. {\em Landau-Zener formulae from transition histories}, in preparation.

\bibitem[BeTe$_3$]{BeTe3} V.\ Betz and S.\ Teufel. {\em Adiabatic transition
histories for Born-Oppenheimer type models}, in preparation.

%\bibitem[BMKNZ]{BMKNZ} A.\ Bohm, A.\ Mostafazadeh, H.\ Koizumi, Q.\ Niu and J.\ Zwanziger.
%{\em The geometric phase in quantum systems}, Texts and Monographs
%in Physics, Springer, Heidelberg, 2003.

\bibitem[Bo]{Bo} J.\ P.\ Boyd. {\em The Devil's Invention: Asymptotics, Superasymptotic and Hyperasymptotic Series}, Acta Applicandae {\bf 56}, 1-98 (1999).


\bibitem[Di]{Di} R.\ B.\ Dingle. {\em Asymptotic Expansions: Their Derivation and Interpretation.} Academic Press (1973).



\bibitem[HaJo]{HaJo} G.\ Hagedorn and A.\ Joye. {\em Time development of
exponentially small non-adiabatic transitions}, Commun.\ Math.\
Phys.\ {\bf 250}, 393--413 (2004).

%\bibitem[He1]{He1} Peter Henrici. {\em Applied and computational analysis, Vol.\ 1}, Wiley (1974).

\bibitem[He]{He} P.\ Henrici. {\em Applied and computational analysis, Vol.\ 2}, Wiley (1977).

\bibitem[Jo]{Jo} A.\ Joye. {\em Non-trivial prefactors in adiabatic
transition probabilities induced by high order complex
degeneracies}, J.\ Phys.\ A {\bf 26}, 6517--6540 (1993).

\bibitem[JKP]{JKP} A.\ Joye, H.\ Kunz and C.-E.\ Pfister. {\em
Exponential decay and geometric aspect of transition probabilities
in the adiabatic limit}, Ann.\ Phys.\ {\bf 208}, 299 (1991).

%\bibitem[JoPf$_1$]{JoPf1} A.\ Joye and C.-E.\ Pfister. {\em
%Exponentially small adiabatic invariant for the Schr\"odinger
%equation},
% Commun.\ Math.\ Phys.\ {\bf 140}, 15--41 (1991).


\bibitem[JoPf]{JoPf} A.\ Joye and C.-E.\ Pfister. {\em
Superadiabatic evolution and adiabatic transition probability
between two nondegenerate levels isolated in the spectrum},
 J.\ Math.\ Phys.\ {\bf 34}, 454--479 (1993).



%\bibitem[Ka]{Ka}  T.\ Kato. \emph{On the adiabatic theorem of
%quantum mechanics}, Phys.\ Soc.\ Jap.\ \textbf{5}, 435--439 (1950).

\bibitem[Ma]{Ma} A.\ Martinez. {\em Precise exponential estimates
in adiabatic theory}, J.\ Math.\ Phys.\ {\bf 35}, 3889--3915 (1994).

%\bibitem[Le]{Le} A.\ Lenard. {\em Adiabatic invariants to all orders}, Ann.\ Phys.\ {\bf 6}, 261--276 (1959).

\bibitem[LiBe]{LiBe} R.\ Lim and M.\ V.\ Berry. {\em Superadiabatic tracking of quantum
evolution}, J.\ Phys.\ A {\bf 24}, 3255--3264 (1991).

%\bibitem[Ne$_1$]{Nen1} G.\ Nenciu. {\em Adiabatic theorem and spectral concentration},
%Commun.\ Math.\ Phys.\ {\bf 82}, 121--135 (1981).


\bibitem[Ne]{Ne}  G.\ Nenciu. \emph{Linear adiabatic theory.
Exponential estimates}, Commun.\ Math.\ Phys.\ \textbf{152},
479--496 (1993).

\bibitem[PST]{PST} G.\ Panati, H.\ Spohn and S.\ Teufel. {\em Space-adiabatic perturbation theory},
Adv.\ Theor.\ Math.\ Phys.\ {\bf 7}, 145--204 (2003).


\bibitem[Sj]{Sj}  J.\ Sj\"{o}strand. \emph{Projecteurs adiabatiques
du point de vue pseudodiff\'{e}rentiel}, C.\ R.\ Acad.\ Sci.\ Paris
S\'{e}r.\ I Math.\ \textbf{317}, 217--220 (1993).

%\bibitem[ReSi]{ReSi2} M.\ Reed and B.\ Simon. {\em Methods of modern mathematical
%physics II}, Academic Press (1975).


\bibitem[Te]{Te} S.\ Teufel. {\em Adiabatic perturbation theory in
quantum dynamics}, Springer Lecture Notes in Mathematics 1821, 2003.

%\bibitem[WiMo]{WiMo} M.\ Wilkinson and M.\ Morgan. {\em
%Nonadiabatic transitions in multilevel systems}, Phys.\ Rev.\ A {\bf
%61}, 062104 (2000).

\end{thebibliography}
\end{document}